\RequirePackage{fix-cm}
\documentclass[smallextended]{svjour3}       
\usepackage{natbib}

\smartqed  
\usepackage{graphicx}
\journalname{Empirical Software Engineering}
\usepackage{amsmath,amssymb,amsfonts}
\usepackage{algorithmic}
\usepackage{textcomp}
\usepackage{rotating}
\usepackage{multirow}
\usepackage{booktabs}
\usepackage{tabularx}
\usepackage{mdframed}
\usepackage{longtable}
\usepackage{caption}
\usepackage{subcaption}

\usepackage{inconsolata}

\usepackage{color}

\definecolor{pblue}{rgb}{0.13,0.13,1}
\definecolor{pgreen}{rgb}{0,0.5,0}
\definecolor{pred}{rgb}{0.9,0,0}
\definecolor{pgrey}{rgb}{0.46,0.45,0.48}

\usepackage{listings}
\usepackage{hyperref}

\begin{document}

\newcommand\setrow[1]{\gdef\rowmac{#1}#1\ignorespaces}
\newcommand\clearrow{\global\let\rowmac\relax}
\clearrow

\title{A Fine-grained Data Set and Analysis of Tangling in Bug Fixing Commits}
\titlerunning{A Fine-grained Data Set and Analysis of Tangling in Bug Fixing Commits}

\author{
Steffen Herbold \and
Alexander Trautsch \and
Benjamin Ledel \and
Alireza Aghamohammadi \and
Taher Ahmed Ghaleb \and
Kuljit Kaur Chahal \and
Tim Bossenmaier \and
Bhaveet Nagaria \and
Philip Makedonski \and
Matin Nili Ahmadabadi \and
Kristof Szabados \and
Helge Spieker \and
Matej Madeja \and
Nathaniel Hoy \and
Valentina Lenarduzzi \and
Shangwen Wang \and
Gema Rodríguez-Pérez \and
Ricardo Colomo-Palacios \and
Roberto Verdecchia \and
Paramvir Singh \and
Yihao Qin \and
Debasish Chakroborti \and
Willard Davis \and
Vijay Walunj \and
Hongjun Wu \and
Diego Marcilio \and
Omar Alam \and
Abdullah Aldaeej \and
Idan Amit \and
Burak Turhan \and
Simon Eismann \and
Anna-Katharina Wickert \and
Ivano Malavolta \and
Mat\'u\v{s} Sul\'ir \and
Fatemeh Fard \and
Austin Z. Henley \and
Stratos Kourtzanidis  \and
Eray Tuzun \and
Christoph Treude \and
Simin Maleki Shamasbi \and
Ivan Pashchenko \and
Marvin Wyrich \and
James Davis \and
Alexander Serebrenik \and
Ella Albrecht \and
Ethem Utku Aktas \and
Daniel Strüber \and
Johannes Erbel}

\institute{
Steffen Herbold\\Institute for Software and Systems Engineering, TU Clausthal, Clausthal-Zellefeld, Germany, Germany\\
\email{steffen.herbold@kit.edu}
\vspace{5pt}\\
Alexander Trautsch\\Institute of Computer Science, University of Goettingen, Goettingen, Germany\\
\email{alexander.trautsch@cs.uni-goettingen.de}
\vspace{5pt}\\
Benjamin Ledel\\Institute for Software and Systems Engineering, TU Clausthal, Clausthal-Zellefeld, Germany, Germany\\
\email{benjamin.ledel@tu-clausthal.de}
\vspace{5pt}\\
Alireza Aghamohammadi\\Department of Computer Engineering, Sharif University of Technology, Tehran, Iran\\
\email{aaghamohammadi@ce.sharif.edu}
\vspace{5pt}\\
Taher Ahmed Ghaleb\\School of Computing, Queen's University, Kingston, Canada\\
\email{taher.ghaleb@queensu.ca}
\vspace{5pt}\\
Kuljit Kaur Chahal, Department of Computer Science, Guru Nanak Dev University, Amritsar, India\\
\email{kuljitchahal.cse@gndu.ac.in}
\vspace{5pt}\\
Tim Bossenmaier\\Karlsruhe Institute of Technology (KIT), Karlsruhe, Germany\\
\email{udeho@student.kit.edu}
\vspace{5pt}\\
Bhaveet Nagaria\\Brunel University London, Uxbridge, United Kingdom\\
\email{bhaveet.nagaria@brunel.ac.uk}
\vspace{5pt}\\
Philip Makedonski \\Institute of Computer Science, University of Goettingen, Germany \\
\email{makedonski@cs.uni-goettingen.de}
\vspace{5pt}\\
Matin Nili Ahmadabadi\\University of Tehran, Tehran, Iran\\
\email{matin\_nili@alumni.ut.ac.ir}
\vspace{5pt}\\
Krist\'of Szabados\\Ericsson Hungary ltd., Budapest, Hungary \\
\email{kristof.szabados@ericsson.com}
\vspace{5pt}\\
Helge Spieker\\Simula Research Laboratory, Fornebu, Norway\\
\email{helge@simula.no}
\vspace{5pt}\\
Matej Madeja\\Technical University of Košice, Košice, Slovakia \\
\email{matej.madeja@tuke.sk}
\vspace{5pt}\\
Nathaniel G. Hoy\\Brunel University London, Uxbridge, United Kingdom\\
\email{nathaniel.hoy2@brunel.ac.uk}
\vspace{5pt}\\
Valentina Lenarduzzi\\LUT University, Finland \\
\email{valentina.lenarduzzi@lut.fi}
\vspace{5pt}\\
Shangwen Wang\\National University of Defense Technology, Changsha, China\\
\email{wangshangwen13@nudt.edu.cn}
\vspace{5pt}\\
Gema Rodríguez-Pérez\\University of British Columbia, Kelowna, Canada\\
\email{gema.rodriguezperez@ubc.ca}
\vspace{5pt}\\
Ricardo Colomo-Palacios\\ \O{}stfold University College, Halden, Norway \\
\email{ricardo.colomo-palacios@hiof.no}
\vspace{5pt}\\
Roberto Verdecchia\\Vrije Universiteit Amsterdam, Amsterdam, The Netherlands\\
\email{r.verdecchia@vu.nl}
\vspace{5pt}\\
Paramvir Singh\\University of Auckland, Auckland, New Zealand\\
\email{p.singh@auckland.ac.nz}
\vspace{5pt}\\
Yihao Qin\\National University of Defense Technology, Changsha, China\\
\email{yihaoqin@nudt.edu.cn}
\vspace{5pt}\\
Debasish Chakroborti\\University of Saskatchewan, Saskatoon, Canada\\
\email{debasish.chakroborti@usask.ca}
\vspace{5pt}\\
Willard Davis\\IBM, Boulder, USA \\
\email{wdavis@us.ibm.com} 
\vspace{5pt}\\
Vijay Walunj\\University of Missouri-Kansas City, Kansas City, USA\\
\email{vbwgh6@umsystem.edu}
\vspace{5pt}\\
Hongjun Wu\\National University of Defense Technology, Changsha, China\\
\email{wuhongjun15@nudt.edu.cn}
\vspace{5pt}\\
Diego Marcilio\\Università della Svizzera italiana, Lugano, Switzerland \\
\email{diego.marcilio@usi.ch}
\vspace{5pt}\\
Omar Alam\\Trent University, Peterborough, Canada\\
\email{omaralam@trentu.ca}
\vspace{5pt}\\
Abdullah Aldaeej\\
University of Maryland Baltimore County, United States and
Imam Abdulrahman Bin Faisal University, Saudi Arabia\\
\email{aldaeej1@umbc.edu}
\vspace{5pt}\\
Idan Amit\\The Hebrew University/Acumen, Jerusalem, Israel \\
\email{idan.amit@mail.huji.ac.il}
\vspace{5pt}\\
Burak Turhan\\University of Oulu, Oulu, Finland and Monash University, Melbourne, Australia \\
\email{burak.turhan@oulu.fi} 
\vspace{5pt}\\
Simon Eismann\\University of Würzburg, Würzburg, Germany \\
\email{simon.eismann@uni-wuerzburg.de} 
\vspace{5pt}\\
Anna-Katharina Wickert\\Technische Universität Darmstadt, Darmstadt, Germany \\
\email{wickert@cs.tu-darmstadt.de}
\vspace{5pt}\\
Ivano Malavolta\\Vrije Universiteit Amsterdam, Amsterdam, The Netherlands \\
\email{i.malavolta@vu.nl}
\vspace{5pt}\\
Mat\'u\v{s} Sul\'ir \\Technical University of Ko\v{s}ice, Ko\v{s}ice, Slovakia \\
\email{matus.sulir@tuke.sk} 
\vspace{5pt}\\
Fatemeh Fard\\University of British Columbia, Kelowna, Canada\\
\email{Fatemeh.fard@ubc.ca}
\vspace{5pt}\\
Austin Z. Henley\\University of Tennessee, Knoxville, USA \\
\email{azh@utk.edu}
\vspace{5pt}\\
Stratos Kourtzanidis\\University of Macedonia, Thessaloniki, Greece \\
\email{ekourtzanidis@uom.edu.gr}  
\vspace{5pt}\\
Eray T\"{u}z\"{u}n\\Department of Computer Engineering, Bilkent University, Ankara, Turkey\\
\email{eraytuzun@cs.bilkent.edu.tr}
\vspace{5pt}\\
Christoph Treude\\University of Melbourne, Melbourne, Australia\\
\email{christoph.treude@unimelb.edu.au}
\vspace{5pt}\\
Simin Maleki Shamasbi\\Independent Researcher, Tehran, Iran\\
\email{simin.maleki@gmail.com}
\vspace{5pt}\\
Ivan Pashchenko \\ University of Trento, Trento, Italy \\
\email{ivan.pashchenko@unitn.it}
\vspace{5pt}\\
Marvin Wyrich\\University of Stuttgart, Stuttgart, Germany \\
\email{marvin.wyrich@iste.uni-stuttgart.de} 
\vspace{5pt}\\
James C. Davis\\Purdue University, West Lafayette, IN, USA\\
\email{davisjam@purdue.edu}
\vspace{5pt}\\
Alexander Serebrenik\\Eindhoven University of Technology, Eindhoven, The Netherlands\\
\email{a.serebrenik@tue.nl}
\vspace{5pt}\\
Ella Albrecht\\Institute of Computer Science, University of Goettingen, Germany\\
\email{ella.albrecht@cs.uni-goettingen.de}
\vspace{5pt}\\
Ethem Utku Aktas \\Softtech Inc., Research and Development Center, 34947 Istanbul, Turkey \\
\email{utku.aktas@softtech.com.tr}
\vspace{5pt}\\
Daniel Strüber\\Radboud University, Nijmegen, Netherlands \\
\email{d.strueber@cs.ru.nl}
\vspace{5pt}\\
Johannes Erbel\\Institute of Computer Science, University of Goettingen, Germany\\
\email{johannes.erbel@cs.uni-goettingen.de}
}

\date{Received: date / Accepted: date}

\maketitle

\begin{abstract}
\textit{Context:} Tangled commits are changes to software that address multiple concerns at once. For researchers interested in bugs, tangled commits mean that they actually study not only bugs, but also other concerns irrelevant for the study of bugs.

\textit{Objective:} We want to improve our understanding of the prevalence of tangling and the types of changes that are tangled within bug fixing commits. 

\textit{Methods:} We use a crowd sourcing approach for manual labeling to validate which changes contribute to bug fixes for each line in bug fixing commits. Each line is labeled by four participants. If at least three participants agree on the same label, we have consensus. 

\textit{Results:} We estimate that between 17\% and 32\% of all changes in bug fixing commits modify the source code to fix the underlying problem. However, when we only consider changes to the production code files this ratio increases to 66\% to 87\%. We find that about 11\% of lines are hard to label leading to active disagreements between participants. Due to confirmed tangling and the uncertainty in our data, we estimate that 3\% to 47\% of data is noisy without manual untangling, depending on the use case.

\textit{Conclusion:} Tangled commits have a high prevalence in bug fixes and can lead to a large amount of noise in the data. Prior research indicates that this noise may alter results. As researchers, we should be skeptics and assume that unvalidated data is likely very noisy, until proven otherwise.

\end{abstract}

\section{Introduction}

Detailed and accurate information about bug fixes is important for many different domains of software engineering research, e.g., program repair~\citep{Gazzola2019}, bug localization~\citep{Mills2018}, and defect prediction~\citep{Hosseini2017a}. Such research suffers from mislabeled data, e.g., because commits are mistakenly identified as bug fixes~\citep{Herzig2013} or because not all changes within a commit are bug fixes~\citep{Herzig2013a}. A common approach to obtain information about bug fixes is to mine
software repositories for bug fixing commits and assume that all changes in the bug fixing commit are part of the bug fix, e.g., with the SZZ algorithm~\citep{Sliwerski2005}. Unfortunately, prior research showed that the reality is more complex. The term \emph{tangled commit}\footnote{\cite{Herzig2013a} actually used the term tangled change. However, we follow \cite{Perez2020} and use the term commit to reference observable groups of changes within version control systems (see Section~\ref{sec:terminology}).} was established by \cite{Herzig2013a} to characterize the problem that commits may address multiple issues, together with the concept of untangling, i.e., the subsequent separation of these concerns. Multiple prior studies established through manual validation that tangled commits naturally occur in code bases. For example, \cite{Herzig2013a}, \cite{Nguyen2013}, \cite{Kirinuki2014}, \cite{Kochar2014}, \cite{Kirinuki2016}, \cite{Wang2019}, and \cite{Mills2020}. Moreover, \cite{Nguyen2013}, \cite{Kochar2014}, \cite{Herzig2016}, and \cite{Mills2020} have independently shown that tangling can have a negative effect on experiments, e.g., due to noise in the training data that reduces model performance as well as due to noise in the test data which significantly affects performance estimates. 

However, we identified four limitations regarding the knowledge on tangling within the current literature.The first and most common limitation of prior work is that it either only considered a sample of commits, or that the authors failed to determine whether the commits were tangled or not for a large proportion of the data.\footnote{Failing to label a proportion of the data also results in a sample, but this sample is highly biased towards changes that are simple to untangle.} Second, the literature considers this problem from different perspectives: most literature considers tangling at the commit level, whereas others break this down to the file-level within commits.  Some prior studies focus on all commits, while others only consider bug fixing commits. Due to these differences, in combination with limited sample sizes, the prior work is unclear regarding the prevalence of tangling.
For example, \cite{Kochar2014} and \cite{Mills2020} found a high prevalence of tangling within file changes, but their estimates for the prevalence of tangling varied substantially with 28\% and 50\% of file changes affected, respectively. Furthermore, how the lower boundary of 15\% tangled commits that \cite{Herzig2013a} estimated relates to the tangling of file changes is also unclear. Third, there is little work on what kind of changes are tangled. \cite{Kirinuki2014} studied the type of code changes that are often tangled with other changes and found that these are mostly logging, checking of pre-conditions, and refactorings. \cite{Nguyen2013} and \cite{Mills2020} provide estimations on the types of tangled commits on a larger scale, but their results contradict each other. For example, \cite{Mills2020} estimate six times more refactorings in tangled commits than \cite{Nguyen2013}. Fourth, these studies are still relatively coarse-grained and report results at the commit and file level. 

Despite the well established research on tangled commits and their impact on research results, their prevalence and content is not yet well understood, neither for commits in general, nor with respect to bug fixing commits. Due to this lack of understanding, we cannot estimate how severe the threat to the validity of experiments due to the tangling is, and how many tools developed based on tangled commits may be negatively affected. 

Our lack of knowledge about tangled commits notwithstanding, researchers need data about bug fixing commits. Currently, researchers rely on three different approaches to mitigate this problem: (i) seeded bugs; (ii) no or heuristic untangling; and (iii) manual untangling. First, we have data with seeded bugs, e.g., the SIR data~\citep{Do2005}, the Siemens data~\citep{Hutchins1994}, and through mutation testing~\citep{Jia2011}. While there is no risk of noise in such data, it is questionable whether the data is representative for real bugs. Moreover, applications like bug localization or defect prediction cannot be evaluated based on seeded bugs. 

The second approach is to assume that commits are either not tangled or that heuristics are able to filter tangling. Examples of such data are ManyBugs~\citep{Goues2015}, Bugs.jar~\citep{Saha2018}, as well as all defect prediction data sets~\citep{Herbold2019}. The advantage of such data sets is that they are relatively easy to collect. The drawback is that the impact of noise due to tangled commits is unclear, even though modern variants of the SZZ algorithm can automatically filter some tangled changes like comments, whitespaces~\citep{Kim2006} or even some refactorings~\citep{Neto2018}.

Third, there are also some data sets that were manually untangled. While the creation of such data is very time consuming, such data sets are the gold standard. They represent real-world defects and do not contain noise due to tangled commits. To the best of our knowledge, there are only very few such data sets, i.e., Defects4J~\citep{Just2014} with Java bugs, BugsJS~\citep{Gyimesi2019} with JavaScript bugs, and the above mentioned data by \cite{Mills2020}.\footnote{The cleaning performed by \cite{Mills2020} is not full untangling of the bug fixes, as all pure additions were also flagged as tangled. While this is certainly helpful for bug localization, as a file that was added as part of the bug fix cannot be localized based on an issue description, this also means that it is unclear which additions are part of the bug fix and which additions are tangled.} Due to the effort required to manually validate changes, the gold standard data sets contain only samples of bugs from each studied project, but no data covers all reported bugs within a project, which limits the potential use cases.\footnote{We note that none of the gold standard data sets actually has the goal to contain data for all bugs of a project. This is likely also not possible due to other requirements on these data sets, e.g., the presence of failing test cases. Thus, the effort is not the only involved factor why these data sets are only samples.}

This article fills this gap in our current knowledge about tangling bug fixing commits. We provide a new large-scale data set that contains validated untangled bug fixes for the complete development history of 23 Java projects and partial data for five further projects. In comparison to prior work, we label all data on a line-level granularity. We have (i) labels for each changed line in a bug fixing commit; (ii) accurate data about which lines contribute to the semantic change of the bug fix; and (iii) the kind of change contained in the other lines, e.g., whether it is a change to tests, a refactoring, or a documentation change. This allows us not only to untangle the bug fix from all other changes, but also gives us valuable insights into the content of bug fixing commits in general and the prevalence of tangling within such commits.

Through our work, we also gain a better understanding of the limitations of manual validation for the untangling of commits. Multiple prior studies indicate that there is a high degree of uncertainty when determining whether a change is tangled or not~\citep{Herzig2013a, Kirinuki2014, Kirinuki2016}. Therefore, we present each commit to four different participants who labeled the data independently from each other. We use the agreement among the participants to gain insights into the uncertainty involved in the labeling, while also finding lines where no consensus was achieved, i.e., that are hard for researchers to classify.

Due to the massive amount of manual effort required for this study, we employed the \emph{research turk} approach~\citep{Herbold2020} to recruit participants for the labeling. The research turk is a means to motivate a large number of researchers to contribute to a common goal, by clearly specifying the complete study together with an open invitation and clear criteria for participation beforehand~\citep{Herbold2020a}. In comparison to surveys or other studies where participants are recruited, participants in the research turk actively contribute to the research project, in our case by labeling data and suggesting improvements of the manuscript. As a result, 45 of the participants we recruited became co-authors of this article. Since this is, to the best of our knowledge, a new way to conduct research projects about software engineering, we also study the effectiveness of recruiting and motivating participants. 

Overall, the contributions of this article are the following. 
\begin{itemize}
    \item The \emph{Line-Labelled Tangled Commits for Java (LLTC4J)} corpus of manually validated bug fixing commits covering 2,328 bugs from 28 projects that were fixed in 3,498 commits that modified 289,904 lines. Each changed line is annotated with the type of change, i.e., whether the change modifies the source code to correct the problem causing the bug, a whitespace or documentation change, a refactoring, a change to a test, a different type of improvement unrelated to the bug fix, or whether the participants could not determine a consensus label. 
    \item Empirical insights into the different kinds of changes within bug fixing commits which indicate that, in practice, most changes in bug fixing commits are not about the actual bug fix, but rather related changes to non-production artifacts such as tests or documentation. 
    \item We introduce the concept of \emph{problematic tangling} for different uses of bug data to understand the noise caused by tangled commits for different research applications.
    \item We find that researchers tend to unintentionally mislabel lines in about 7.9\% of the cases. Moreover, we found that identifying refactorings and other unrelated changes seems to be challenging, which is shown through 14.3\% of lines without agreement in production code files, most of which are due to a disagreement whether a change is part of the bug fix or unrelated. 
    \item This is the first use of the research turk method and we showed that this is an effective research method for large-scale studies that could not be accomplished otherwise. 
\end{itemize}

The remainder of this article is structured as follows. We discuss the related work in Section~\ref{sec:related-work}. We proceed with the description of the research protocol in Section~\ref{sec:research-protocol}. We present and discuss the results for our research questions in Section~\ref{sec:rq1} and Section~\ref{sec:rq2}. We report the threats to the validity of our work in Section~\ref{sec:threats}. Finally, we conclude in Section~\ref{sec:conclusion}.

\section{Related Work}
\label{sec:related-work}

We focus the discussion of the related work on the manual untangling of commits. Other aspects, such as automated untangling algorithms~\citep[e.g.,][]{Kreutzer2016, Partachi2020}, the separation of concerns into multiple commits~\citep[e.g.,][]{Ryo2018, Yamashita2020}, the tangling of features with each other~\citep{Struder2020}, the identification of bug fixing or inducing commits~\citep[e.g.,][]{Perez2020}, or the characterization of commits in general~\citep[e.g.,][]{Hindle2008}, are out of scope. 

\subsection{Magnitude of Tangling}

The tangling of commits is an important issue, especially for researchers working with software repository data, that was first characterized by \cite{Kawrykow2011}. They analyzed how often non-essential commits occur in commit histories, i.e., commits that do not modify the logic of the source code and found that up to 15.5\% of all commits are non-essential. Due to the focus on non-essential commits, the work by \cite{Kawrykow2011} only provides a limited picture on tangled commits in general, as tangling also occurs if logical commits for multiple concerns are mixed within a single commit. 

The term \emph{tangled change} (commit) was first used in the context of repository mining by \cite{Herzig2013a} (extended in \cite{Herzig2016}\footnote{In the following, we cite the original paper, as the extension did not provide further evidence regarding the tangling.}). The term tangling itself was already coined earlier in the context of the separation of concerns~\citep[e.g.,][]{Kiczales1997}. \cite{Herzig2013a} studied the commits of five Java projects over a period of at least 50 months and tried to manually classify which of the commits were tangled, i.e., addressed multiple concerns. Unfortunately, they could not determine if the commits are tangled for 2,498 of the 3,047 bug fixing commits in their study. Of the bug fixing commits they could classify, they found that 298 were tangled and 251 were not tangled. Because they could not label large amounts of the commits, they estimate that at least 15\% of the bug fixing commits are tangled. 

\cite{Nguyen2013} also studied tangling in bug fixing commits, but used the term mixed-purpose fixing commits. They studied 1296 bug fixing commits from eight projects and identified 297 tangled commits, i.e., 22\% of commits that are tangled. A strength of the study is that the authors did not only label the tangled commits, but also identified which file changes in the commit were tangled. Moreover, \cite{Nguyen2013} also studied which types of changes are tangled and found that 4.9\% were due to unrelated improvements, 1.1\% due to refactorings, 1.8\% for formatting issues, and 3.5\% for documentation changes unrelated to the bug fix.\footnote{These percentages are not reported in the paper. We calculated them from their Table III. We combined enhancement and annotations into unrelated improvements, to be in line with our work.} We note that it is unclear how many of the commits are affected by multiple types of tangling and, moreover, that the types of tangling were analyzed for only about half of the commits. Overall, this is among the most comprehensive studies on this topic in the literature. However, there are two important issues that are not covered by \cite{Nguyen2013}. First, they only explored if commits and files are tangled, but not how much of a commit or file is tangled. Second, they do not differentiate between problematic tangling and benign tangling (see Section~\ref{sec:impact}), and, most importantly, between tangling in production files and tangling in other files. Thus, we cannot estimate how strongly tangling affects different research purposes.

\cite{Kochar2014} investigated the tangling of a random sample of the bug fixing commits for 100 bugs. Since they were interested in the implications of tangling on bug localization, they focused on finding out how many of the changed files actually contributed to the correction of the bug, and how many file changes are tangled, i.e., did not contribute to the bug fix. They found that 358 out of 498 file changes were corrective, the remaining 140 changes, i.e., 28\% of all file changes in bug fixing commits were tangled. 

\cite{Kirinuki2014} and \cite{Kirinuki2016} are a pair of studies with a similar approach: they used an automated heuristic to identify commits that may be tangled and then manually validated if the identified commits are really tangled. \cite{Kirinuki2014} identified 63 out of 2,000 commits as possibly tangled and found through manual validation that 27 of these commits were tangled and 23 were not tangled. For the remaining 13 commits, they could not decide if the commits are tangled. \cite{Kirinuki2016} identified 39 out of 1,000 commits as potentially tangled and found through manual validation that 21 of these commits were tangled, while 7 were not tangled. For the remaining 11 commits, they could not decide. A notable aspect of the work by \cite{Kirinuki2014} is that they also analyzed what kind of changes were tangled. They found that tangled changes were mostly due to logging, condition checking, or refactorings. We further note that these studies should not be seen as an indication of low prevalence of tangled commits, just because only 48 out of 3,000 commits were found to be tangled, since only 102 commits were manually validated. These commits are not from a representative random sample, but rather from a sample that was selected such that the commits should be tangled according to a heuristic. Since we have no knowledge about how good the heuristic is at identifying tangled commits, it is unclear how many of the 2,898 commits that were not manually validated are also tangled. 

\cite{Tao2015} investigated the impact of tangled commits on code review. They manually investigated 453 commits and found that 78 of these commits were tangled in order to determine ground truth data for the evaluation of their code review tool. From the description within the study, it is unclear if the sample of commits is randomly selected, if these are all commits in the four target projects within the study time frame that changed multiple lines, or if a different sampling strategy was employed. Therefore, we cannot conclude how the prevalence of tangling in about 17\% of the commits generalizes beyond the sample. 

\cite{Wang2019} also manually untangled commits. However, they had the goal to create a data set for the evaluation of an automated algorithm for the untangling of commits. They achieved this by identifying 50 commits that they were sure were tangled commits, e.g., because they referenced multiple issues in the commit message. They hired eight graduate students to perform the untangling in pairs. Thus, they employed an approach for the untangling that is also based on the crowd sourcing of work, but within a much more closely controlled setting and with only two participants per commit instead of four participants. However, because \cite{Wang2019} only studied tangled commits, no conclusions regarding the prevalence of the tangling can be drawn based on their results. 

\cite{Mills2020} extends \cite{Mills2018} and contains the most comprehensive analysis of tangled commits to date. They manually untangled the changes for 803 bugs from fifteen different Java projects. They found that only 1,154 of the 2,311 changes of files within bug fixing commits contributed to bug fixes, i.e., a prevalence of 50\% of tangling. Moreover, they provide insights into what kinds of changes are tangled: 31\% of tangled changes are due to tests, 9\% due to refactorings, and 8\% due to documentation. They also report that 47\% of the tangled changes are due to adding source code. Unfortunately, \cite{Mills2020} flagged all additions of code as tangled changes. While removing added lines makes sense for the use case of bug localization, this also means that additions that are actually the correction of a bug would be wrongly considered to be tangled changes. Therefore, the actual estimation of prevalence of tangled changes is between 32\% and 50\%, depending on the number of pure additions that are actually tangled. We note that the findings by \cite{Mills2020} are not in line with prior work, i.e., the ratio of tangled file changes is larger than the estimation by \cite{Kochar2014} and the percentages for types of changes are different from the estimations by \cite{Nguyen2013}. 

\cite{Dias2015} used a different approach to get insights into the tangling of commits: they integrated an interface for the untangling of commits directly within an IDE and asked the developers to untangle commits when committing their local changes to the version control system. They approached the problem of tangled commits by grouping related changes into clusters. Unfortunately, the focus of the study is on the developer acceptance, as well as on the changes that developers made to the automatically presented clusters. It is unclear if the clusters are fully untangled, and also if all related changes are within the same cluster. Consequently, we cannot reliably estimate the prevalence or contents of tangled commits based on their work. 

\subsection{Data Sets}

Finally, there are data sets of bugs where the data was untangled manually, but where the focus was only on getting the untangled bug fixing commits, not on the analysis of the tangling. \cite{Just2014} did this for the Defects4J data and \cite{Gyimesi2019} for the BugsJS data. While these data sets do not contain any noise, they have several limitations that we overcome in our work. First, we do not restrict the bug fixes but allow all bugs from a project. Defects4J and BugsJS both only allow bugs that are fixed in a single commit and also require that a test that fails without the bug fix was added as part of the bug fix. While bugs that fulfill these criteria were manually untangled for Defects4J, BugJS has additional limitations. BugJS requires that bug fixes touch at most three files, modify at most 50 lines of code, and do not contain any refactorings or other improvements unrelated to the bug fix. Thus, BugsJS is rather a sample of bugs that are fixed in untangled commits than a sample of bugs that was manually untangled. While these data sets are gold standard untangled data sets, they are not suitable to study tangling. Moreover, since both data sets only contain samples, they are not suitable for research that requires all bugs of a specific release or within a certain time frame of a project. Therefore, our data is suitable for more kinds of research than Defects4J and BugsJS, and because our focus is not solely on the creation of the data set, but also on the understanding of tangling within bug fixing commits, we also provide a major contribution to the knowledge about tangled commits. 

\subsection{Research Gap}

Overall, there is a large body of work on tangling, but studies are limited in sampling strategies, inability to label all data, or because the focus was on different aspects. Thus, we cannot form conclusive estimates, neither regarding the types of tangled changes, nor regarding the general prevalence of tangling within bug fixing commits.

\section{Research Protocol}
\label{sec:research-protocol}

Within this section, we discuss the research protocol, i.e., the research strategy we pre-registered~\citep{Herbold2020a} to study our research questions and hypotheses. The description and section structure of our research protocol is closely aligned with the pre-registration, but contains small deviations, e.g., regarding the sampling strategy. All deviations are described as part of this section and summarized in Section~\ref{sec:deviations}.

\subsection{Terminology}
\label{sec:terminology}

We use the term ``the principal investigators'' to refer to the authors of the registered report, ``the manuscript'' to refer to this article that resulted from the study, and ``the participants'' to refer to researchers and practitioners who collaborated on this project, many of whom are now co-authors of this article. 

Moreover, we use the term ``commit'' to refer to observable changes within the version control system, which is in line with the terminology proposed by \cite{Perez2020}. We use the term ``change'' to refer to the actual modifications within commits, i.e., what happens in each line that is modified, added, or deleted as part of a commit. 

\subsection{Research Questions and Hypotheses}
\label{sec:rqs}

Our research is driven by two research questions for which we derived three hypotheses. The first research question is the following. 

\noindent
\textbf{RQ1}: What percentage of changed lines in bug fixing commits contributes to the bug fix and can we identify these changes?

We want to get a detailed understanding of both the prevalence of tangling, as well as what kind of changes are tangled within bug fixing commits. When we speak of contributing to the bug fix, we mean a direct contribution to the correction of the logical error. We derived two hypotheses related to this research question. 

\begin{itemize}
  \item[\textbf{H1}] Fewer than 40\% of changed lines in bug fixing commits contribute to the bug fix.
  \item[\textbf{H2}] A label is a \emph{consensus label} when at least three participants agree on it. Participants fail to achieve a consensus label on at least 10.5\% of lines.\footnote{In the pre-registered protocol, we use the number of at least 16\%. However, this was a mistake from the calculation of the binomial distribution, where we used a wrong value $n=5$ instead of $n=4$. This was the result of our initial plan to use five participants per commit, which we later revised to four participants without updating the calculation and the hypothesis.}
\end{itemize}

We derived hypothesis H1 from the work by \cite{Mills2018}, who found that 496 out of 1344 changes to files in bug fixing commits contributed to bug fixes.\footnote{The journal extension by \cite{Mills2020} was not published when we formulated our hypotheses as part of the pre-registration of our research protocol in January 2020.} We derived our expectation from \cite{Mills2018} instead of \cite{Herzig2013} due to the large degree of uncertainty due to unlabeled data in the work by \cite{Herzig2013}. 
We derived H2 based on the assumption that there is a 10\% chance that participants misclassify a line, even if they have the required knowledge for correct classification. We are not aware of any evidence regarding the probability of random mistakes in similar tasks and, therefore, used our intuition to estimate this number. Assuming a binomial distribution $B(k|p, n)$ with the probability of random mislabels $p=0.1$ and $n=4$ participants that label each commit, we do not get a consensus for $k \geq 2$ participants randomly mislabeling the line, i.e.,
\begin{equation}
\sum_{k=2}^4 B(k| 0.1, 4) = \sum_{k=2}^4 {\binom{4}{k}} 0.1^k \cdot 0.9^{4-k} = 0.0523  
\end{equation}
Thus, if we observe 10.5\% of lines without consensus, this would be twice more than expected given the assumption of 10\% random errors, indicating that lack of consensus is not only due to random errors. We augment the analysis of this hypothesis with a survey among participants, asking them how often they were unsure about the labels. 

The second research question regards our crowd working approach to organize the manual labor required for our study. 

\noindent
\textbf{RQ2:} Can gamification motivate researchers to contribute to collaborative research projects?
\begin{itemize}
  \item[\textbf{H3}] The leaderboard motivates researchers to label more than the minimally required 200 commits.
\end{itemize}

We derived H3 from prior evidence that gamification~\citep{Werbach2012} is an efficient method for the motivation of computer scientists, e.g., as demonstrated on Stack Overflow~\citep{Grant2013}. Participants can view a nightly updated leaderboard, both to check their progress, as well as where they would currently be ranked in the author list. We believe that this has a positive effect on the amount of participation, which we observe through larger numbers of commits labeled than minimally required for co-authorship. We augment the analysis of this hypothesis with a survey among participants, asking them if they were motivated by the leaderboard and the prospect of being listed earlier in the author list. 

\subsection{Materials}

This study covers bug fixing commits that we re-use from prior work (see Section~\ref{sec:bugs}). We use SmartSHARK to process all data~\citep{Trautsch2018}. We extended SmartSHARK with the capability to annotate the changes within commits. This allowed us to manually validate which lines in a bug fixing commit are contributing to the semantic changes for fixing the bugs~\citep{Trautsch2020}.

\subsection{Variables}
\label{sec:variables}

We now state the variables we use as foundation for the construct of the analysis we conduct. The measurement of these variables is described in Section~\ref{sec:consensus} and their analysis is described in Section~\ref{sec:analysis-phase}.

For bug fixing commits, we measure the following variables as percentages of lines with respect to the total number of changed lines in the commit.
\begin{itemize}
  \item Percentage contributing to bug fixes.
  \item Percentage of whitespace changes.
  \item Percentage of documentation changes.
  \item Percentage of refactorings.
  \item Percentage of changes to tests.
  \item Percentage of unrelated improvements.
  \item Percentage where no consensus was achieved (see Section~\ref{sec:consensus}).
\end{itemize}

We provide additional results regarding the lines without consensus to understand what the reasons for not achieving consensus are. These results are an extension of our registered protocol. However, we think that details regarding potential limitations of our capabilities as researchers are important for the discussion of research question RQ1. Concretely, we consider the following cases:

\begin{itemize}
    \item In the registration, we planned to consider whitespace and documentation changes within a single variable. Now, we use separate variables for both. Our reason for this extension is to enable a better understanding of how many changes are purely cosmetic without affecting functionality (whitespace) and distinguish this from changes that modify the documentation. We note that we also used the term ``comment'' instead of ``documentation'' within the registration. Since all comments (e.g., in code) are a form of documentation, but not all documentation (e.g., sample code) is a comment, we believe that this terminology is clearer.
    \item Lines where all labels are either test, documentation, or whitespace. We use this case, because our labeling tool allows labeling of all lines in a file with the same label with a single click and our tutorial describes how to do this for a test file. This leads to differences between how participants approached the labeling of test files: some participants always use the button to label the whole test file as test, other participants used a more fine-grained approach and also labeled whitespace and documentation changes within test files.
    \item Lines that were not labeled as bug fix by any participant. For these lines, we have consensus that this is not part of the bug fix, i.e., for our key concern. Disagreements may be possible, e.g., if some participants identified a refactoring, while others marked this as unrelated improvement. 
\end{itemize}

A second deviation from our pre-registered protocol is that we present the results for the consensus for two different views on the data:
\begin{itemize}
    \item all changes, i.e., as specified in registration; and
    \item only changes in Java source code files that contain production code, i.e., Java files excluding changes to tests or examples.
\end{itemize}

\begin{table}
    \begin{tabular}{l|p{9cm}}
        \textbf{File type} & \textbf{Regular expression}\\
        \hline
        Test & \verb!(^|\/)(test|tests|test_long_running|testing|legacy-tests! \verb!|testdata|test-framework|derbyTesting|unitTests|java\/stubs! \verb!|test-lib|src\/it|src-lib-test|src-test|tests-src|test-cactus! \verb!|test-data|test-deprecated|src_unitTests|test-tools|! \verb!gateway-test-release-utils|gateway-test-ldap|nifi-mock)\/!\\
        Documentation & \verb!(^|\/)(doc|docs|example|examples|sample|samples|demo|tutorial! \verb!|helloworld|userguide|showcase|SafeDemo)\/!\\
        Other & \verb!(^|\/)(_site|auxiliary-builds|gen-java|external! \verb!|nifi-external)\/!\\
    \end{tabular}
    \caption{Regular expressions for excluding non-production code files. These regular expressions are valid for a superset of our data and were manually validated as part of prior work~\cite{Trautsch2020a}.}
    \label{tbl:regex}
\end{table}

Within the pre-registered protocol, we do not distinguish between all changes and changes to production code. We now realize that both views are important. The view of all changes is required to understand what is actually part of bug fixing commits. The view on Java production files is important, because changes to non-production code can be easily determined automatically as not contributing to the bug fix, e.g., using regular expression matching based on the file ending ``.java'' and the file path to exclude folders that contain tests and example code. The view on Java production files enables us to estimate the amount of noise that cannot be automatically removed. Within this study, we used the regular expressions shown in Table~\ref{tbl:regex} to identify non-production code. We note that we have some projects which also provide web applications (e.g., JSP Wiki), that also have production code in other languages than Java, e.g., JavaScript. Thus, we slightly underestimate the amount of production code, because these lines are only counted in the overall view, and not in the production code view.

Additionally, we measure variables related to the crowd working.
\begin{itemize}
  \item Number of commits labeled per participant.
  \item Percentage of correctly labeled lines per participant, i.e., lines where the label of the participant agrees with the consensus.
\end{itemize}
We collected this data using nightly snapshots of the number of commits that each participant labeled, i.e., a time series per participant.

We also conducted an anonymous survey among participants who labeled at least 200 commits with a single question to gain insights into the difficulty of labeling tangled changes in commits for the evaluation of RQ1.
\begin{itemize}
  \item \textit{Q1:} Please estimate the percentage of lines in which you were unsure about the label you assigned.
  \item \textit{A1:} One of the following categories: 0\%--10\%, 11\%--20\%, 21\%--30\%, 31\%--40\%, 41\%--50\%, 51--60\%, 61\%--70\%, 71\%--80\%, 81\%--90\%, 91\%--100\%.
\end{itemize}

Finally, we asked all participants who labeled at least 250 commits a second question to gain insights into the effect of the gamification:
\begin{itemize}
  \item \textit{Q2:} Would you have labeled more than 200 commits, if the authors would have been ordered randomly instead of by the number of commits labeled?
  \item \textit{A2:} One of the following categories: Yes, No, Unsure.
\end{itemize}

\subsection{Subjects}

This study has two kinds of subjects: bugs for which the lines contributing to the fix are manually validated and the participants in the labeling who were recruited using the research turk approach.

\subsubsection{Bugs}
\label{sec:bugs}

We use manually validated bug fixes. For this, we harness previously manually validated issue types similar to \cite{Herzig2013} and validated trace links between commits and issues for 39 Apache projects~\citep{Herbold2019}. The focus of this data set is the Java programming language. The usage of manually validated data allows us to work from a foundation of ground truth and avoids noise in our results caused by the inclusion of commits that are not fixing bugs. Overall, there are 10,878 validated bug fixing commits for 6,533 fixed bugs in the data set.

Prior studies that manually validated commits limited the scope to bugs that were fixed in a single commit, and commits in which the bug was the only referenced issue~\citep[e.g.,][]{Just2014, Gyimesi2019, Mills2020}. We broaden this scope in our study and also allow issues that were addressed by multiple commits, as long as the commits only fixed a single bug. This is the case for 2,283 issues which were addressed in 6,393 commits. Overall, we include 6,279 bugs fixed in 10,389 commits in this study. The remaining 254 bugs are excluded, because the validation would have to cover the additional aspect of differentiating between multiple bug fixes. Otherwise, it would be unclear to which bug(s) the change in a line would contribute, making the labels ambiguous. 

\cite{Herbold2019} used a purposive sampling strategy for the determination of projects. They selected only projects from the Apache Software Foundation, which is known for the high quality of the developed software, the many contributors both from the open source community and from the industry, as well as the high standards of their development processes, especially with respect to issue tracking~\cite{Bissyande2013}. Moreover, the 39 projects cover different kinds of applications, including build systems (ant-ivy), web applications (e.g., jspwiki), database frameworks (e.g., calcite), big data processing tools (e.g., kylin), and general purpose libraries (commons). Thus, our sample of bug fixes should be representative for a large proportion of Java software. Additionally, \cite{Herbold2019} defined criteria on project size and activity to exclude very small or inactive projects. Thus, while the sample is not randomized, this purposive sampling should ensure that our sample is representative for mature Java software with well-defined development processes in line with the discussion of representativeness by \cite{Baltes2020}.

\subsubsection{Participants}
\label{sec:participants}

In order to only allow participants that have a sufficient amount of programming experience, each participant must fulfill one of the following criteria: 1) an undergraduate degree majoring in computer science or a closely related subject; or 2) at least one year of programming experience in Java, demonstrated either by industrial programming experience using Java or through contributions to Java open source projects.

Participants were regularly recruited, e.g., by advertising during virtual conferences, within social media, or by asking participants to invite colleagues. Interested researchers and practitioners signed up for this study via an email to the first author, who then checked if the participants are eligible. Upon registration, participants received guidance on how to proceed with the labeling of commits (see Appendix~\ref{app:instructions}). Participants became co-authors of this manuscript if
\begin{enumerate}
    \item they manually labeled at least 200 commits;
    \item their labels agree with the consensus (Section~\ref{sec:consensus}) for at least 70\% of the labeled lines;
    \item they contributed to the manuscript by helping to review and improve the draft, including the understanding that they take full responsibility for all reported results and that they can be held accountable with respect to the correctness and integrity of the work; and
    \item they were not involved in the review or decision of acceptance of the registered report. 
\end{enumerate}

The first criterion guarantees that each co-author provided a significant contribution to the analysis of the bug fixing commits. The second criterion ensures that participants carefully labeled the data, while still allowing for disagreements. Only participants who fulfill the first two criteria received the manuscript for review. The third criterion ensures that all co-authors agree with the reported results and the related responsibility and ethical accountability. The fourth criterion avoids conflicts of interest.\footnote{Because the review of the registered report was blinded, the fulfillment of this criterion is checked by the editors of the Empirical Software Engineering journal.}

\subsection{Execution Plan}

The execution of this research project was divided into two phases: the data collection phase and the analysis phase. 

\subsubsection{Data Collection Phase}
\label{sec:consensus}

\begin{figure}
\centering
\includegraphics[width=\textwidth]{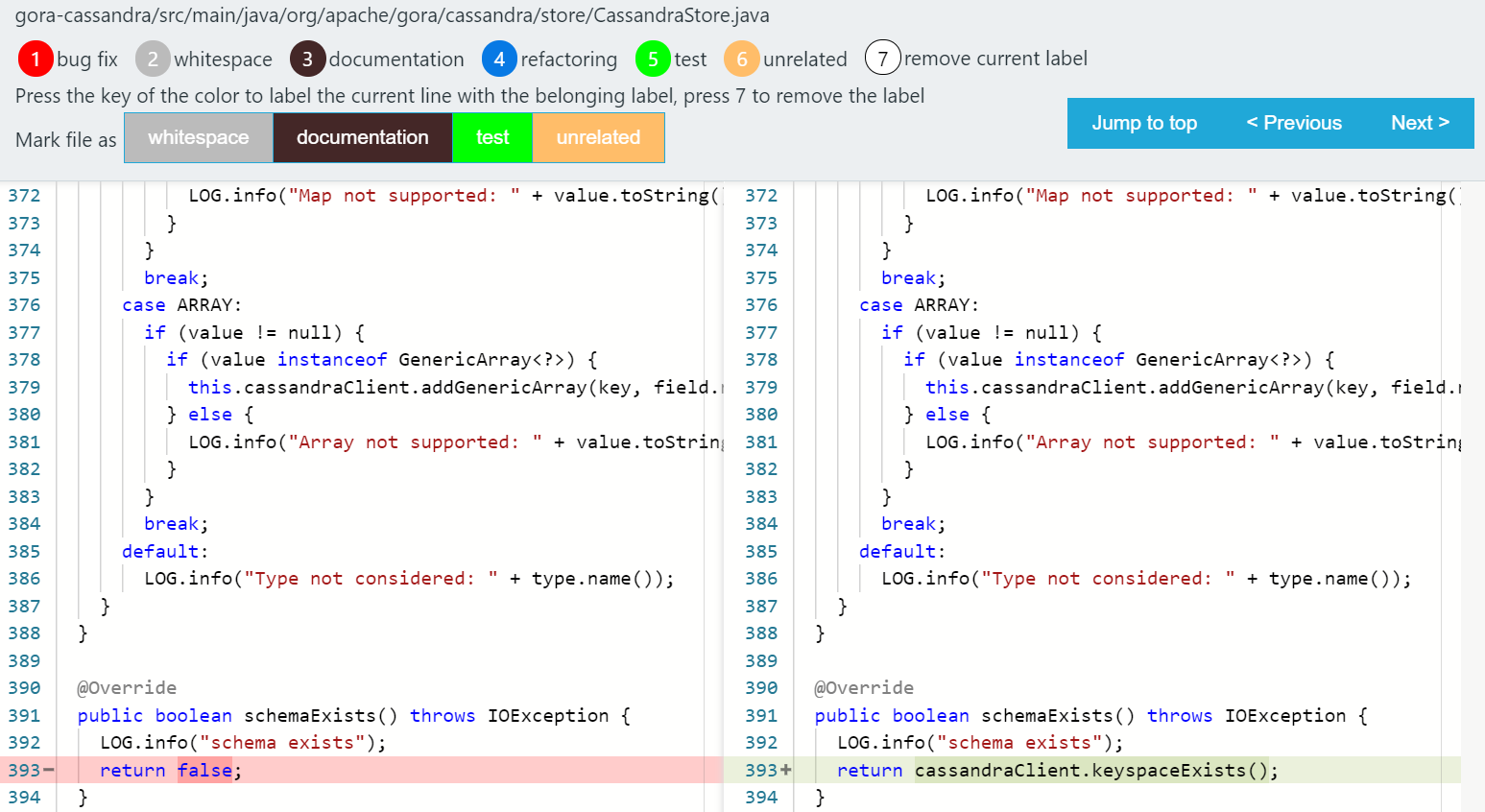}
\caption{Screenshot of VisualSHARK~\citep{Trautsch2020} that was used for the labeling of the data.}
\label{fig:screenshot}
\end{figure}

The primary goal of this study is to gain insights into which changed lines contribute to bug fixing commits and which additional activities are tangled with the correction of the bug. Participants were shown the textual differences of the source code for each bug with all related bug fixing commits. The participants then assigned one of the following labels to all changed lines:
\begin{itemize}
    \item contributes to the bug fix;
    \item only changes to whitespaces;
    \item documentation change;
    \item refactoring;
    \item change to tests; and
    \item unrelated improvement not required for the bug fix. 
\end{itemize}

Figure~\ref{fig:screenshot} shows a screenshot of the web application that we used for labeling. The web application ensured that all lines were labeled, i.e., participants could not submit incomplete labels for a commit. The web application further supported the participants with the following convenience functions: 
\begin{itemize}
    \item Buttons to mark all changes in a file with the same label. This could, e.g., be used to mark all lines in a test file with a single click as test change. 
    \item Heuristic pre-labeling of lines as documentation changes using regular expressions.
    \item Heuristic pre-labeling of lines with only whitespace changes.
    \item Heuristic pre-labeling of lines as refactoring by automatically marking changed lines as refactorings, in case they were detected as refactorings by the RefactoringMiner 1.0~\citep{Tsantalis2018}.
\end{itemize}

All participants were instructed not to trust the pre-labels and check if these labels are correct. Moreover, we did not require differentiation between whitespaces and documentation/test changes in files that were not production code files, e.g., within test code or documentation files such as the project change log. 

Each commit was shown to four participants. Consensus is achieved if at least three participants agree on the same label. If this is not the case, no consensus for the line is achieved, i.e., the participants could not clearly identify which type of change a line is. 

The data collection phase started on May 16th, 2020 and ended on October 14th, 2020, registration already finished on September 30th, 2020.\footnote{This timeframe is a deviation from the registration protocol that was necessary due to the Covid-19 pandemic.} The participants started by watching a tutorial video\footnote{\url{https://www.youtube.com/watch?v=VWvDlq4lQC0}} and then labeling the same five commits that are shown in the tutorial themselves to get to know the system and to avoid mislabels due to usability problems. 

Participants could always check their progress, as well as the general progress for all projects and the number of commits labeled by other participants in the leaderboard. However, due to the computational effort, the leaderboard did not provide a live view of the data, but was only updated once every day. All names in the leaderboard were anonymized, except the name of the currently logged in participant and the names of the principal investigators. The leaderboard is also part of a gamification element of our study, as participants can see how they rank in relation to others and may try to improve their ranks by labeling more commits.

The participants are allowed to decide for which project they want to perform the labeling. The bugs are then randomly selected from all bugs of that project, for which we do not yet have labels by four participants. We choose this approach over randomly sampling from all projects to allow participants to gain experience in a project, which may improve both the labeling speed and the quality of the results. Participants must finish labeling each bug they are shown before the next bug can be drawn. We decided for this for two reasons. First, skipping bugs could reduce the validity of the results for our second research question, i.e., how good we actually are at labeling bug fixes at this level of granularity, because the sample could be skewed towards simpler bug fixes. Second, this could lead to cherry picking, i.e., participants could skip bugs until they find particularly easy bugs. This would be unfair for the other participants.The drawback of this approach is that participants are forced to label bugs, even in case they are unsure. However, we believe that this drawback is countered by our consensus labeling that requires agreement of three participants: even if all participants are unsure about a commit, if three come to the same result, it is unlikely that they are wrong. 

\subsubsection{Analysis Phase}
\label{sec:analysis-phase}

The analysis took place after the data collection phase was finished on October 14th, 2020. In this phase, the principal investigators conducted the analysis as described in Section~\ref{sec:analysis-plan}. The principal investigators also informed all participants of their consensus ratio. All participants who met the first two criteria for co-authorship received a draft of the manuscript for review. Due to the number of participants, this was conducted in multiple batches. In between, the principal investigators improved the manuscript based on the comments of the participants. All participants who reviewed the draft were added to the list of authors. Those who failed to provide a review were added to the acknowledgements, unless they specifically requested not to be added. Finally, all co-authors received a copy of the final draft one week in advance of the submission for their consideration. 

\subsubsection{Data Correction Phase}

Due to a bug in the data collection software that we only found after the data collection was finished, we required an additional phase for the correction of data. The bug led to possibly corrupt data in case line numbers in the added and deleted code overlapped, e.g., single line modifications. We computed all lines that were possibly affected by this bug and found that 28,827 lines possibly contained corrupt labels, i.e., about 10\% of our data. We asked all co-authors to correct their data through manual inspection by relabeling all lines that could have been corrupted. For this, we used the same tooling as for the initial labeling, with the difference that all labels that were not affected by the bugs were already set, only the lines affected by the bug needed to be relabeled.\footnote{Details can be found in the tutorial video: \url{https://www.youtube.com/watch?v=Kf6wVoo32Mc}} The data correction phase took place from November 25th, 2020 until January 17th, 2021. We deleted all data that was not corrected by January 18th which resulted in 679 less issues for which the labeling has finished. We invited all co-authors to re-label these issues between January 18th and January 21th. Through this, the data for 636 was finished again, but we still lost the data for 43 issues as the result of this bug. The changes to the results were very minor and the analysis of the results was not affected. However, we note that the analysis of consensus ratios and participation (Section~\ref{sec:participation}) was done prior to the correction phase. The bug affected only a small fraction of lines that should only have a negligible impact on the individual consensus ratios, as data by all participants was affected equally. We validated this intuition by comparing the consensus ratios of participants who corrected their data before and after the correction and found that there were no relevant changes.  Since some participants could not participate in the data correction and we had to remove some labelled commits, the number of labeled commits per author could now be below 200. This altered the results due to outliers, caused by single very large commits. This represents an unavoidable trade-off arising from the need to fix the potentially corrupt data.

\subsection{Analysis Plan}
\label{sec:analysis-plan}

The analysis of data consists of four aspects: the contribution to bug fixes, the capability to label bug fixing commits, the effectiveness of gamification, and the confidence level of our statistical analysis. 

\subsubsection{Contributions to Bug Fixes}

We used the Shapiro-Wilk test~\citep{Shapiro1965} to determine if the nine variables related to defects are normally distributed. Since the data is not normal, we report the median, median absolute deviation (MAD), and an estimation for the confidence interval of the median based on the approach proposed by~\cite{Campbell1988}. We reject H1 if the upper bound of the confidence interval of the median lines contributing to the bug fix in all code is greater than 40\%. Within the discussion, we provide further insights about the expectations on lines within bug fixing commits based on all results, especially also within changes to production code files. 

\subsubsection{Capability to Label Bug Fixing Commits}
\label{sec:mistakes}

\begin{table}
\centering
\begin{tabular}{cl}
$\kappa$ & \textbf{Interpretation} \\
\hline \hline
$<$0 & Poor agreement \\
0.01 – 0.20 & Slight agreement \\
0.21 – 0.40 & Fair agreement \\
0.41 – 0.60 & Moderate agreement \\
0.61 – 0.80 & Substantial agreement \\
0.81 – 1.00 & Almost perfect agreement \\
\end{tabular}
\caption{Interpretation of Fleiss' $\kappa$ according to \cite{Landis1977}.}
\label{tbl:kappa}
\end{table}

We use the confidence interval for the number of lines without consensus for this evaluation. We reject H2 if the lower bound of the confidence interval of the median number of lines without consensus is less than 10.5\%. Additionally, we report Fleiss' $\kappa$~\citep{Fleiss1971} to estimate the reliability of the consensus, which is defined as
\begin{equation}
\kappa = \frac{\bar{P}-\bar{P}_e}{1-\bar{P}_e}
\end{equation}
where $\bar{P}$ is the mean agreement of the participants per line and $\bar{P}_e$ is the sum of the squared proportions of the label assignments. We use the table from  \cite{Landis1977} for the interpretation of $\kappa$ (see Table~\ref{tbl:kappa}).  

Additionally, we estimate the probability of random mistakes to better understand if the lines without consensus can be explained by random mislabels, i.e., mislabels that are the consequence of unintentional mistakes by participants. If we assume that all lines with consensus are labeled correctly, we can use the \emph{minority votes} in those lines to estimate the probability of random mislabels. Specifically, we have a minority vote if three participants agreed on one label, and one participant selected a different label. We assume that random mislabels follow a binomial distribution (see Section~\ref{sec:rqs}) to estimate the probability that a single participant randomly mislabels a bug fixing line. Following \cite{Brown2001}, we use the approach from \cite{Agresti1998} to estimate the probability of a mislabel $p$, because we have a large sample size. Therefore, we estimate
\begin{equation}
p = \frac{n_1+\frac{1}{2}z_\alpha^2}{n+z_\alpha^2}
\end{equation}
as the probability of a mislabel of a participant with $n_1$ the number of minority votes in lines with consensus, $n$ the total number of individual labels in lines with consensus, and $z_\alpha$ the $1-\frac{1}{2}\alpha$ quantile of the standard normal distribution, with $\alpha$ the confidence level. We get the confidence interval for $p$ as
\begin{equation}
p \pm z_\alpha \sqrt{\frac{p\cdot(1-p)}{n+z_\alpha^2}}.
\end{equation}

We estimate the overall probabilities of errors, as well as the probabilities of errors in production files for the different label types to get insights into the distribution of errors. Moreover, we can use the estimated probabilities of random errors to determine how many lines without consensus are expected. If $n_{total}$ is the number of lines, we can expect that there are
\begin{equation}
\begin{split}
n_{none} &= n_{total} \cdot B(k=2|p, n=4) \\
&= n_{total} \cdot {\binom{4}{2}} p^2\cdot (1-p)^2 \\
&= n_{total} \cdot 6 \cdot p^2 \cdot (1-p)^2
\end{split}
\end{equation}
lines without consensus under the assumption that they are due to random mistakes. If we observe more lines without consensus, this is a strong indicator that this is not a random effect, but due to actual disagreements between participants. 

We note that the calculation of the probability of mislabels and the number of expected non-consensus lines was not described in the pre-registered protocol. However, since the approach to model random mistakes as binomial distribution was already used to derive H2 as part of the registration, we believe that this is rather the reporting of an additional detail based on an already established concept from the registration and not a substantial deviation from our protocol.

In addition to the data about the line labels, we use the result of the survey among participants regarding their perceived certainty rates to give further insights into the limitations of the participants to conduct the task of manually labeling lines within commits. We report the histogram of the answers given by the participants and discuss how the perceived difficulty relates to the actual consensus that was achieved.

\subsubsection{Effectiveness of Gamification}

We evaluate the effectiveness of the gamification element of the crowd working by considering the number of commits per participant. Concretely, we use a histogram of the total number of commits per participant. The histogram tells us whether there are participants who have more than 200 commits, including how many commits were actually labeled. Moreover, we create line plots for the number of commits over days, with one line per participant that has more than 200 commits. This line plot is a visualization of the evolution of the prospective ranking of the author list. If the gamification is efficient, we should observe two behavioral patterns: 1) that participants stopped labeling right after they gained a certain rank; and 2) that participants restarted labeling after a break to increase their rank. The first observation would be in line with the results from \cite{Anderson2013} regarding user behavior after achieving badges on Stack Overflow. We cannot cite direct prior evidence for our second conjecture, other than that we believe that participants who were interested in gaining a certain rank, would also check if they still occupy the rank and then act on this. 

We combine the indications from the line plot with the answer to the survey question Q2. If we can triangulate from the line plot and the survey that the gamification was effective for at least 10\% of the overall participants, we believe that having this gamification element is worthwhile and we fail to reject H3. If less than 10\% were motivated by the gamification element, this means that future studies could not necessarily expect a positive benefit, due to the small percentage of participants that were motivated. 
We additionally quantify the effect of the gamification by estimating the additional effort that participants invested measured in the number of commits labeled greater than 200 for the subset of participants where the gamification seems to have made a difference. 

\subsubsection{Confidence Level}

We compute all confidence intervals such that we have a family-wise confidence level of 95\%. We need to adjust the confidence level for the calculation of the confidence intervals, due to the number of intervals we determine. Specifically, we determine $2\cdot 9 = 18$ confidence intervals for the ratios of line labels within commits (see Section~\ref{sec:variables}) and 27 confidence intervals for our estimation of the probability of random mistakes. Due to the large amount of data we have available, we decided for an extremely conservative approach for the adjustment of the confidence level (see Section~\ref{sec:mistakes}). We use Bonferroni correction~\citep{Dunnett1955} for all $18+27=45$ confidence intervals at once, even though we could possibly consider these as separate families. Consequently, we use a confidence level of $1-\frac{0.05}{45} = 0.99\bar{8}$ for all confidence interval calculations.\footnote{The pre-registration only contained correction for six confidence intervals. This increased because we provide a more detailed view on labels without consensus and differentiate between all changes and changes to production code files, and because the calculation of probabilities for mistakes was not mentioned in the registration.} 

\subsection{Summary of Deviations from Pre-Registration}
\label{sec:deviations}

We deviated from the pre-registered research protocol in several points, mostly through the expansion on details. 

\begin{itemize}
\item The time frame of the labeling shifted to May 16th--October 14th. Additionally, we had to correct a part of the data due to a bug between November 25th and January 21st.
\item We updated \textbf{H2} with a threshold of 10.5\% of lines, due to a wrong calculation in the registration (see footnote 6).
\item We consider the subset of mislabels on changes to production code files, as well as mislabels with respect to all changes. 
\item We have additional details because we distinguish between whitespace and documentation lines.
\item We have additional analysis for lines without consensus to differentiate between different reasons for no consensus.
\item We extend the analysis with an estimation of the probability of random mislabels, instead of only checking the percentage of lines without consensus.
\end{itemize}

\begin{table}[t]
\centering
\begin{tabular}{llrr}
\textbf{Project} & \textbf{Timeframe} & \textbf{\#Bugs} & \textbf{\#Commits} \\
\hline\hline
ant-ivy & 2005-06-16 -- 2018-02-13 & 404 / 404 & 547 / 547 \\
\textit{archiva} & 2005-11-23 -- 2018-07-25 & 3 / 278 & 4 / 509 \\
commons-bcel & 2001-10-29 -- 2019-03-12 & 33 / 33 & 52 / 52 \\
commons-beanutils & 2001-03-27 -- 2018-11-15 & 47 / 47 & 60 / 60 \\
commons-codec & 2003-04-25 -- 2018-11-15 & 27 / 27 & 58 / 58 \\
commons-collections & 2001-04-14 -- 2018-11-15 & 48 / 48 & 93 / 93 \\
commons-compress & 2003-11-23 -- 2018-11-15 & 119 / 119 & 205 / 205 \\
commons-configuration & 2003-12-23 -- 2018-11-15 & 140 / 140 & 253 / 253 \\
commons-dbcp & 2001-04-14 -- 2019-03-12 & 57 / 57 & 89 / 89 \\
commons-digester & 2001-05-03 -- 2018-11-16 & 17 / 17 & 26 / 26 \\
commons-io & 2002-01-25 -- 2018-11-16 & 71 / 72 & 115 / 125 \\
commons-jcs & 2002-04-07 -- 2018-11-16 & 58 / 58 & 73 / 73 \\
commons-lang & 2002-07-19 -- 2018-10-10 & 147 / 147 & 225 / 225 \\
commons-math & 2003-05-12 -- 2018-02-15 & 234 / 234 & 391 / 391 \\
commons-net & 2002-04-03 -- 2018-11-14 & 127 / 127 & 176 / 176 \\
commons-scxml & 2005-08-17 -- 2018-11-16 & 46 / 46 & 67 / 67 \\
commons-validator & 2002-01-06 -- 2018-11-19 & 57 / 57 & 75 / 75 \\
commons-vfs & 2002-07-16 -- 2018-11-19 & 94 / 94 & 118 / 118 \\
\textit{deltaspike} & 2011-12-22 -- 2018-08-02 & 6 / 146 & 8 / 219 \\
\textit{eagle} & 2015-10-16 -- 2019-01-29 & 2 / 111 & 2 / 121 \\
giraph & 2010-10-29 -- 2018-11-21 & 140 / 140 & 146 / 146 \\
gora & 2010-10-08 -- 2019-04-10 & 56 / 56 & 98 / 98 \\
\textit{jspwiki} & 2001-07-06 -- 2019-01-11 & 1 / 144 & 1 / 205 \\
opennlp & 2008-09-28 -- 2018-06-18 & 106 / 106 & 151 / 151 \\
parquet-mr & 2012-08-31 -- 2018-07-12 & 83 / 83 & 119 / 119 \\
santuario-java & 2001-09-28 -- 2019-04-11 & 49 / 49 & 95 / 95 \\
\textit{systemml} & 2012-01-11 -- 2018-08-20 & 6 / 279 & 6 / 314 \\
wss4j & 2004-02-13 -- 2018-07-13 & 150 / 150 & 245 / 245 \\
\hline
\textit{Total} && 2328 / 3269 & 3498 / 4855 \\
\hline
\end{tabular}
\caption{Statistics about amounts of labeled data per project, i.e., data for which we have labels by four participants and can compute consensus. The columns \#Bugs and \#Commits list the completed data and the total data available in the time frame. The five projects marked as italic are incomplete, because we did not have enough participants to label all data.}
\label{tbl:project-stats}
\end{table}

\section{Experiments for RQ1}
\label{sec:rq1}

We now present our results and discuss their implications for RQ1 on the tangling of changes within commits.

\subsection{Results for RQ1}

In this section, we first present the data demographics of the study, e.g., the number of participants, and the amount of data that was labeled. We then present the results of the labeling. All labeled data of the LLTC4J corpus and the analysis scripts we used to calculate our results can be found online in our replication package.\footnote{\url{https://github.com/sherbold/replication-kit-2020-line-validation}\\We will move the replication kit to a long-term archive on Zenodo in case of acceptance of this manuscript.}

\subsubsection{Data Demographics}

Of 79 participants registered for this study, 15 participants dropped out without performing any labeling. The remaining 64 participants labeled data. The participants individually labeled 17,656 commits. This resulted in 1,389 commits labeled by one participant, 683 commits labeled by two participants, 303 commits that were labeled by three participants, 3,498 commits labeled by four participants, and five commits that were part of the tutorial were labeled by all participants. Table~\ref{tbl:project-stats} summarizes the completed data for each project. Thus, we have validated all bugs for 23 projects and incomplete data about bugs for five projects. We have a value of Fleiss' $\kappa=0.67$, which indicates that we have substantial agreement among the participants.  

\begin{table}
\centering
\begin{tabular}{lrrrrrr}
\textbf{Label} & \multicolumn{2}{c}{\textbf{All Changes}} & \multicolumn{2}{c}{\textbf{Production Code}} &  \multicolumn{2}{c}{\textbf{Other Code}} \\
\hline\hline
Bug fix & 72774 & (25.1\%) & 71343 & (49.2\%) & 361 & (0.3\%) \\
Test & 114765 & (39.6\%) & 8 & (0.0\%) & 102126 & (91.3\%) \\
Documentation & 40456 & (14.0\%) & 31472 & (21.7\%) & 749 & (0.7\%) \\
Refactoring & 5297 & (1.8\%) & 5294 & (3.7\%) & 3 & (0.0\%) \\
Unrelated Improvement & 1361 & (0.5\%) & 824 & (0.6\%) & 11 & (0.0\%) \\
Whitespace & 11909 & (4.1\%) & 10771 & (7.4\%) & 781 & (0.7\%) \\
\hline
Test/Doc/Whitespace & 4454 & (1.5\%) & 0 & (0.0\%) & 4454 & (4.0\%) \\
No Bug fix & 13052 & (4.5\%) & 4429 & (3.1\%) & 1754 & (1.6\%) \\
No Consensus & 25836 & (8.9\%) & 20722 & (14.3\%) & 1587 & (1.4\%) \\
\hline
\emph{Total} & 289904 && 144863 && 111826 \\
\end{tabular}
\caption{Statistics of assigned line labels over all data. Production code refers all Java files that we did not determine to be part of the test suite or the examples. Other code refers to all other Java files. The labels above the line are for at least three participants selecting the same label. The labels below the line do not have consensus, but are the different categories for lines without consensus we established in Section~\ref{sec:variables}. The eight lines labeled as test in production code are due to two test files within Apache Commons Math that were misplaced in the production code folder.}
\label{tbl:results-overall-lines}
\end{table}

\subsubsection{Content of Bug Fixing Commits}

\begin{figure}
\begin{subfigure}{\textwidth}
\begin{tabular}{lrrrrrrr}
\textbf{Label} & \multicolumn{3}{c}{\textbf{Overall}} & \multicolumn{4}{c}{\textbf{Production Code}} \\
& Med. & MAD & CI & Med. & MAD & CI & $>0$ \\
\hline\hline
Bug fix & 25.0 & 34.8 & [22.2, 29.2] & 75.0 & 37.1 & [70.6, 79.2] & 95.5 \\
Test & 13.0 & 19.3 & [0.0, 23.4] & 0.0 & 0.0 & [0.0, 0.0] & 0.1 \\
Documentation & 10.2 & 15.1 & [7.8, 12.5] & 0.0 & 0.0 & [0.0, 5.1] & 49.4 \\
Refactoring & 0.0 & 0.0 & [0.0, 0.0] & 0.0 & 0.0 & [0.0, 0.0] & 7.8 \\
Unrelated Impr. & 0.0 & 0.0 & [0.0, 0.0] & 0.0 & 0.0 & [0.0, 0.0] & 2.9 \\
Whitespace & 0.0 & 0.0 & [0.0, 0.0] & 0.0 & 0.0 & [0.0, 1.7] & 46.9 \\
\hline
Test/Doc/Whites. & 0.0 & 0.0 & [0.0, 0.0] & 0.0 & 0.0 & [0.0, 0.0] & 0.0 \\
No Bug fix & 0.0 & 0.0 & [0.0, 0.0] & 0.0 & 0.0 & [0.0, 0.0] & 5.8 \\
No Consensus & 0.0 & 0.0 & [0.0, 0.0] & 0.0 & 0.0 & [0.0, 0.0] & 29.7 \\
\hline
\end{tabular}
\caption{The median (Med.), the median absolute deviation (MAD), and the confidence interval of the median (CI) of the percentages of changes within commits. The column $>0$ shows the ratio of commits that have production code file changes of the label type. The many zeros are the consequence of the fact that many commits do not contain any changes that are not of type bug fix, test, or documentation.}
\end{subfigure}
\begin{subfigure}{\textwidth}
\includegraphics[width=\textwidth]{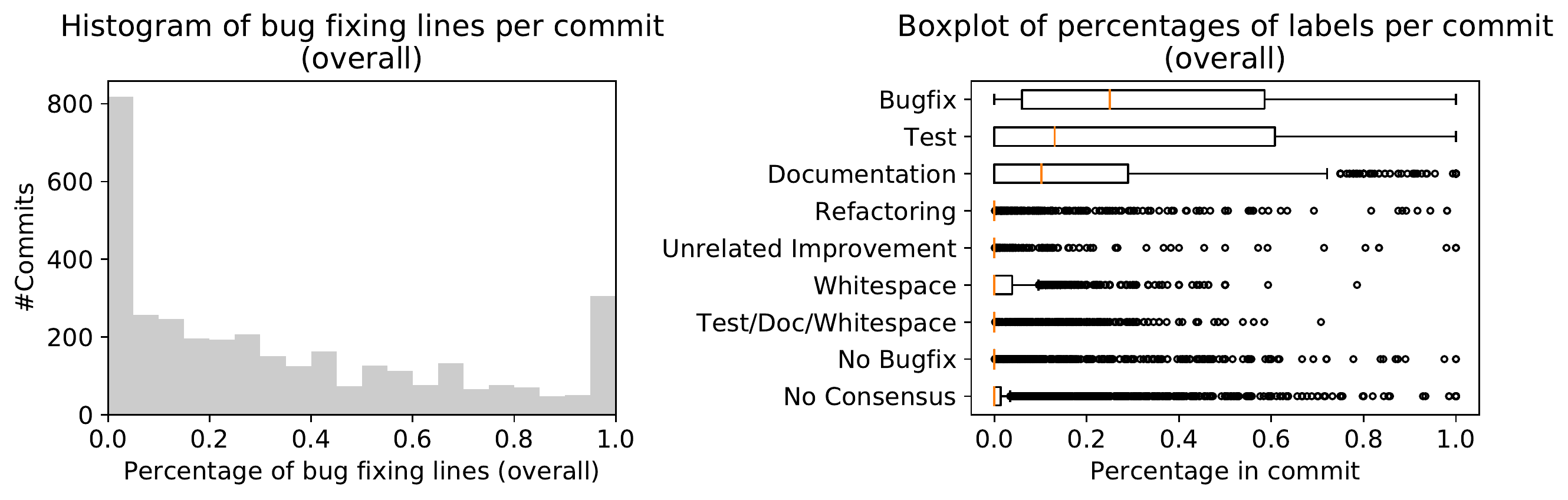}
\caption{Plots for the distribution of lines within commits for all file changes.}
\end{subfigure}
\begin{subfigure}{\textwidth}
\includegraphics[width=\textwidth]{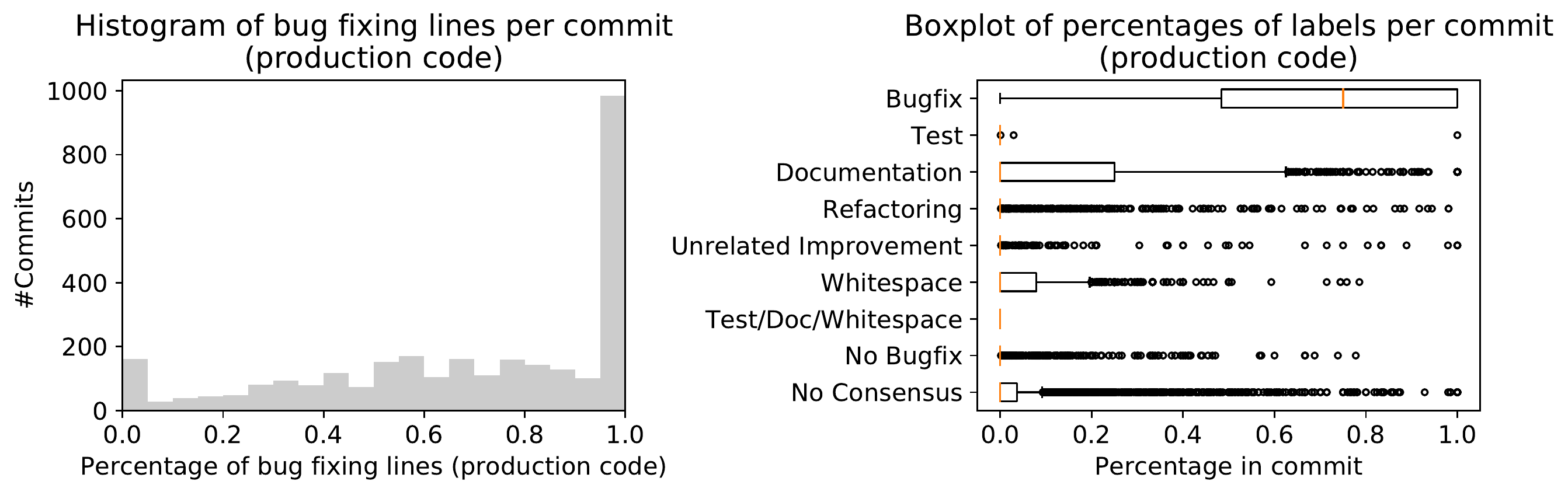}
\caption{Plots for the distribution of lines within commits for changes to production code.}
\end{subfigure}
\caption{Data about changes within commits.}
\label{fig:results-commit-level}
\end{figure}

Table~\ref{tbl:results-overall-lines} summarizes the overall results of the commit labeling. Overall, 289,904 lines were changed as part of the bug fixing commits. Only 25.1\% of the changes were part of bug fixes. The majority of changed lines were modifications of tests with 39.6\%. Documentation accounts for another 14.0\% of the changes. For 8.9\% of the lines there was no consensus among the participants and at least one participant labeled the line as bug fix. For an additional 1.5\% of lines, the participants marked the line as either documentation, whitespace change or test change, but did not achieve consensus. We believe this is the result of different labeling strategies for test files (see Section~\ref{sec:variables}). For 4.5\% of the lines no participant selected bug fix, but at least one participant selected refactoring or unrelated improvement. When we investigated these lines, we found that the majority of cases were due to different labeling strategies: some participants labeled updates to test data as unrelated improvements, others labeled them as tests. How this affected production code files is discussed separately in Section~\ref{sec:disagreements}.

256,689 of the changed lines were Java code, with 144,863 lines in production code files and 11,826 lines in other code files. The other code is almost exclusively test code. Within the production code files, 49.2\% of the changed lines contributed to bug fixes and 21.7\% were documentation. Refactorings and unrelated improvements only represent 4.3\% of the lines. In 14.3\% of the lines, the participants did not achieve consensus with at least one participant labeling the line as bug fix. 

Figure~\ref{fig:results-commit-level} summarizes the results of labeling per commit, i.e., the percentages of each label in bug fixing commits. We found that a median of 25.0\% of all changed lines contribute to a bug fix. When we restrict this to production code files, this ratio increases to 75.0\%. We note that while the median for all changes is roughly similar to the overall percentage of lines that are bug fixing, this is not the case for production code files. The median of 75.0\% per commit is much higher than the 49.2\% of all production lines that are bug fixing. The histograms provide evidence regarding the reason for this effect. With respect to all changed lines, Figure~\ref{fig:results-commit-level}(b) shows that there are many commits with a relatively small percentage of bug fixing lines close to zero, i.e., we observe a peak on the left side of the histogram. When we focus the analysis on the production code files, Figure~\ref{fig:results-commit-level}(c) shows that we instead observe that there are many commits with a large percentage of bug fixing lines (close to 100\%), i.e., we observe a peak on the right side of the histogram, but still a long tail of lower percentages. This tail of lower percentages influences the ratio of lines more strongly in the median per commit, because the ratio is not robust against outliers. 

The most common change to be tangled with bug fixes are test changes and documentation changes with a median of 13.0\% and 10.2\% of lines per commit, respectively. When we restrict the analysis to production code files, all medians other than bug fix drop to zero. For test changes, this is expected because they are, by definition, not in production code files. For other changes, this is due to the extremeness of the data which makes the statistical analysis of most label percentages within commits difficult. What we observe is that while many commits are not pure bug fixes, the type of changes differs between commits, which leads to commits having percentages of exactly zero for all labels other than bug fix. This leads to extremely skewed data, as the median, MAD, and CI often become exactly zero such that the non-zero values are -- from a statistical point of view -- outliers. The last column of the table in Figure~\ref{fig:results-commit-level}(a) shows for how many commits the values are greater than zero. We can also use the boxplots in Figure~\ref{fig:results-commit-level}(b) and (c) to gain insights into the ratios of lines, which we analyze through the outliers. 

The boxplots reveal that documentation changes are common in production code files, the upper quartile is at around 30\% of changed lines in a commit. Unrelated improvements are tangled with 2.9\% of the commits and are usually less than about 50\% of the changes, refactorings are tangled with 7.8\% of the commits and usually less than about 60\% of the changes. Whitespace changes are more common, i.e., 46.9\% of the commits are tangled with some formatting. However, the ratio is usually below 40\% and there are no commits that contain only whitespace changes, i.e., pure reformatting of code.  The distribution of lines without consensus shows that while we have full consensus for 71.3\% of the commits, the consensus ratios for the remaining commits are distributed over the complete range up to no consensus at all. 

As a side note, we also found that the pre-labeling of lines with the RefactoringMiner was not always correct. Sometimes logical changes were marked as refactoring, e.g., because side effects were ignored when variables were extracted or code was reordered. Overall, 21.6\% of the 23,682 lines marked by RefactoringMiner have a consensus label of bug fix. However, the focus of our study is not the evaluation of RefactoringMiner and we further note that we used RefactoringMiner 1.0~\citep{Tsantalis2018} and not the recently released version 2.0~\citep{Tsantalis2020}, which may have resolved some of these issues.

\begin{figure}
\centering
\includegraphics[width=0.5\textwidth]{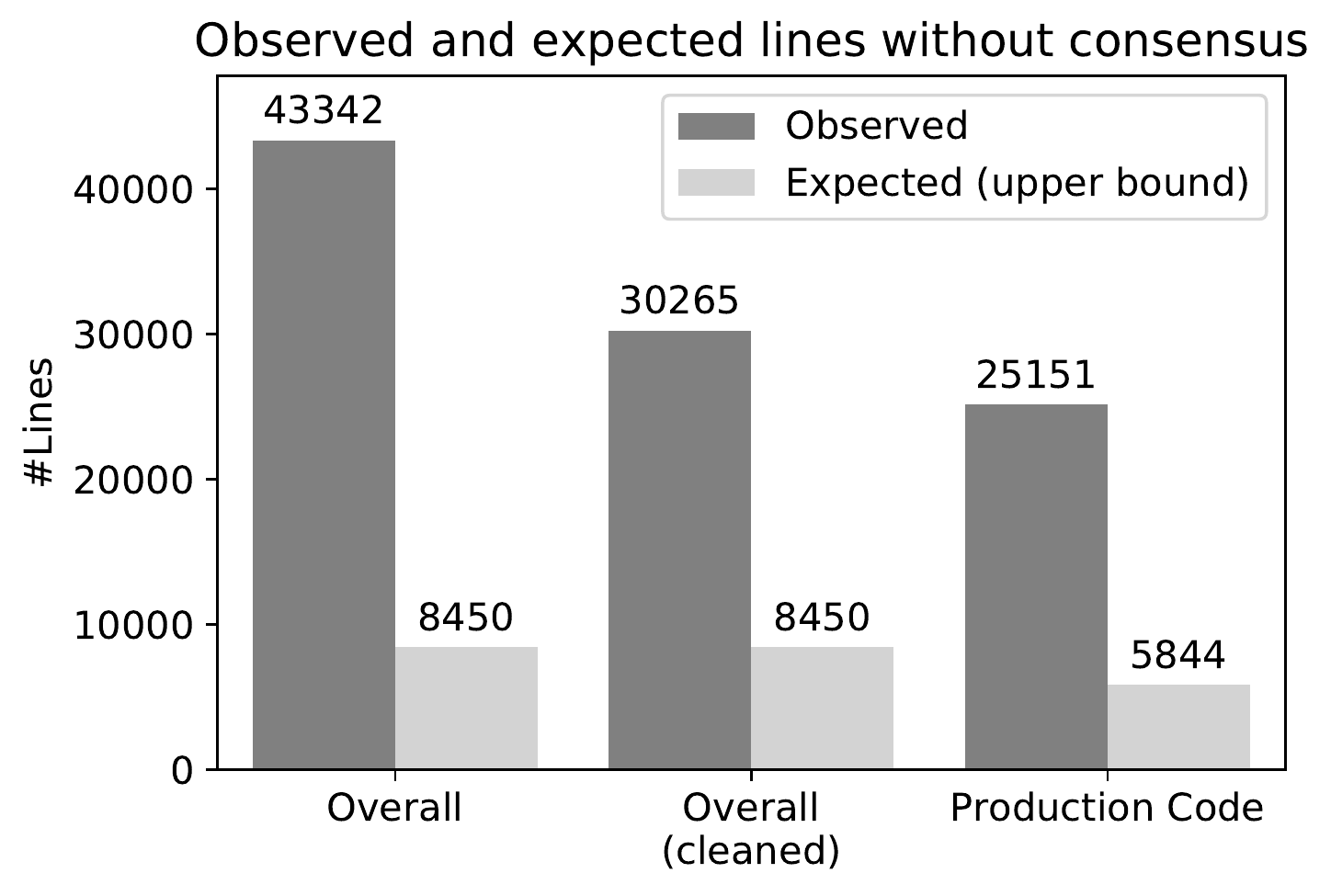}
\caption{Number of expected lines without consensus versus the actual lines without consensus. Overall (cleaned) counts lines with the labels Test/Doc/Whitespace and No Bug fix as consensus.}
\label{fig:exp-mislabels}
\end{figure}

\subsubsection{Analysis of Disagreements}
\label{sec:disagreements}

Based on the minority votes in lines with consensus, we estimate the probability of random mistakes in all changes as $7.5\% \pm 0.0$, when we restrict this to production code files we estimate the probability as $9.0\% \pm 0.0$. We note that the confidence intervals are extremely small, due to the very large number of lines in our data. Figure~\ref{fig:exp-mislabels} shows the expected number of lines without consensus given these probabilities versus the observed lines without consensus. For all changes, we additionally report a cleaned version of the observed data. With the cleaned version, we take the test/doc/whitespace and no bug fix labels into account, i.e., lines where there is consensus that the line is not part of the bug fix.  The results indicate that there are more lines without consensus than could be expected under the assumption that all mislabels in our data are the result of random mislabels. 

Table~\ref{tbl:mislabels-per-type} provides additional details regarding the expectation of the mislabels per type in production code files. The data in this table is modeled using a more fine-grained view on random mistakes. This view models the probabilities for all labels separately, both with respect to expected mislabels, as well as the mislabel that occurs. The estimated probabilities are reported in Table~\ref{tbl:prob-mislabels} in the appendix. Using the probabilities, we calculated the expected number of two random mistakes for each label type, based on the distribution of consensus lines. Table~\ref{tbl:mislabels-per-type} confirms that we do not only observe more lines without consensus, but also more pairs of mislabels than could be expected given a random distribution of mistakes for all labels. However, the mislabels are more often of type bug fix, unrelated improvement, or refactoring than of the other label types. Table~\ref{tbl:no-cons-lines} in the appendix shows a full resolution of the individual labels we observed in lines without consensus and confirms that disagreements between participants whether a line is part of a bug fix, an unrelated improvement, or a refactoring, are the main driver of lines without consensus in production code files.

\begin{table}
\centering
\begin{tabular}{lrrrrrr}
& Bug fix & Doc. & Refactoring & Unrelated & Whitespace & Test \\
\hline
Bug fix & - & 1 & 508 & 705 & 23 & 2 \\
Documentation & 189 & - & 11 & 47 & 17 & 2 \\
Refactoring & 446 & 2 & - & 35 & 3 & 0 \\
Unrelated & 110 & 0 & 2 & - & 1 & 0 \\
Whitespace & 103 & 1 & 49 & 34 & - & 1 \\
\hline
Total Expected: & 847 & 3 & 570 & 821 & 42 & 5 \\
Total Observed: & 12837 & 3987 & 6700 & 8251 & 3951 & 252 \\
\hline
\end{tabular}
\caption{A more detailed resolution of the number of expected mislabels per label type. The rows represent the correct labels, the columns represent the number of two expected mislabels of that type. The sum of the observed values is not equal to the sum of the observed lines without consensus, because it is possible that a line without consensus has two labels of two types each and we cannot know which one is the mislabel. Hence, we must count these lines twice here. The size of the confidence intervals is less than one line in each case, which is why we report only the upper bound of the confidence intervals, instead of the confidence intervals themselves. There is no row for test, because there are no correct test labels in production code files.}
\label{tbl:mislabels-per-type}
\end{table}

\begin{figure}
\centering
\includegraphics[width=0.5\textwidth]{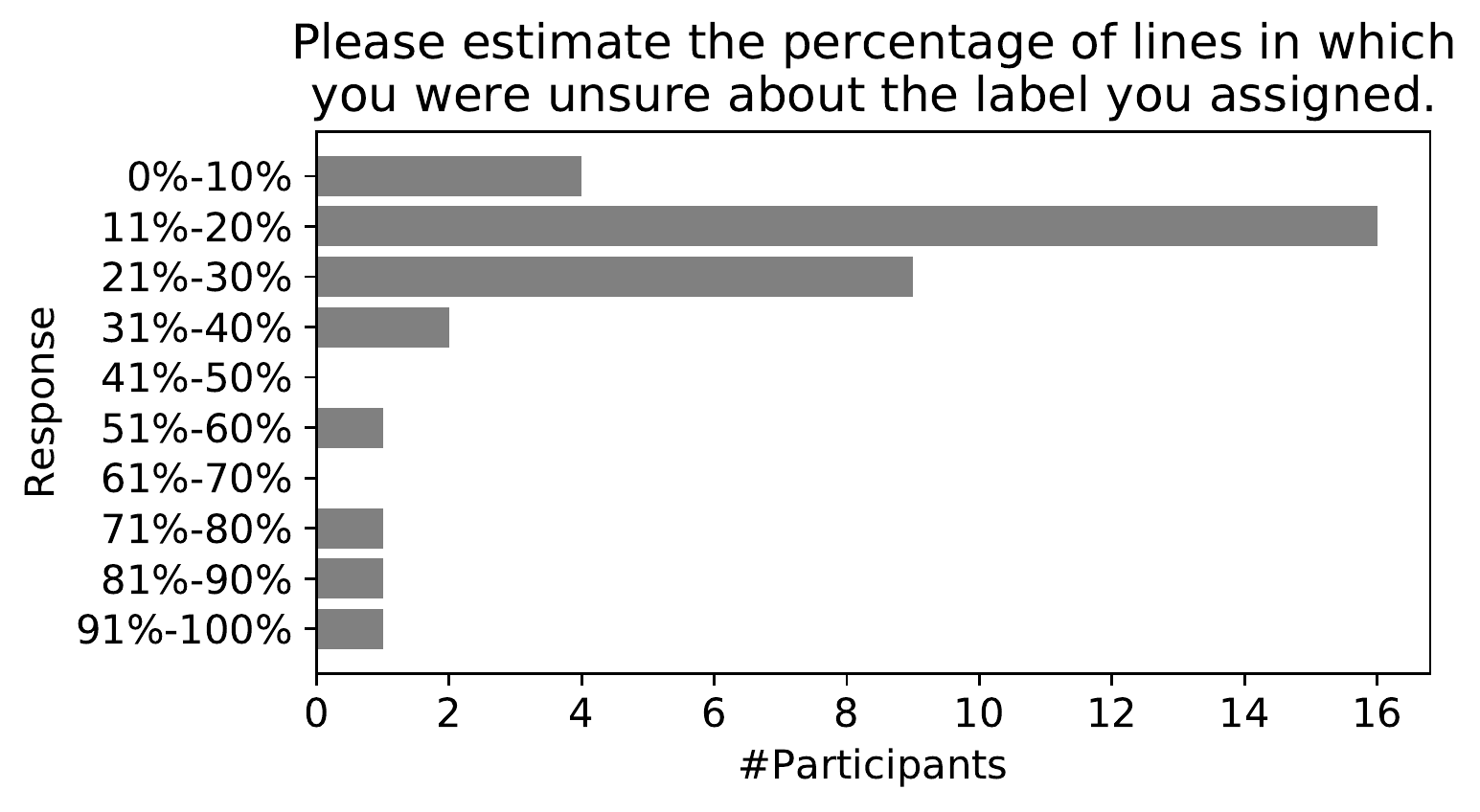}
\caption{Answers of the participants regarding their labeling confidence. Participants were not informed about the distribution of consensus ratios of participants or their individual consensus ratio before answering the survey to avoid bias. 35 out of 48 participants with at least 200 commits answered this question. }
\label{fig:unsure}
\end{figure}

In addition to the view on disagreements through the labels, we also asked the participants that labeled more than 200 commits how confident they were in the correctness of their labels. Figure~\ref{fig:unsure} shows the results of this survey. 55\% of the participants estimate that they were not sure in 11\%--20\% of lines, another 20\% estimate 21\%--30\% of lines. Only one participant estimates a higher uncertainty of 51\%--60\%. This indicates that participants have a good intuition about the difficulty of the labeling. If we consider that we have about 8\% random mistakes and 12\% lines without consensus, this indicates that about 20\% of the lines should have been problematic for participants. We note that there are also three that selected between 71\%--100\% of lines. While we cannot know for sure, we believe that these three participants misread the question, which is in line with up to 10\% of participants found by~\cite{Podsakoff2003}.

\subsection{Discussion of RQ1}
\label{sec:discussion}

We now discuss our results for RQ1 with respect to our hypotheses to gain insights into the research question, put our results in the context of related work, and identify consequences for researchers who analyze bugs. 

\subsubsection{Prevalence of Tangled Commits}

Tangling is common in bug fixing commits. We estimate that the average number of bug fixing lines per commit is between 22\%  and 38\%. The 22\% of lines are due to the lower bound of the confidence interval for the median bug fixing lines per commit. The 38\% of lines assume that all 8.9\% of the lines without consensus would be bug fixing and that the median would be at the upper bound of the confidence interval. However, we observe that a large part of the tangling is not within production code files, but rather due to changes in documentation and test files. Only half of the changes within bug fixing commits are to production code files. We estimate that the average number of bug fixing lines in production code files per commit is between 60\% and 93\%. 

\begin{mdframed}
We fail to reject the hypothesis H1 that less than 40\% of changes in bug fixing commits contribute to the bug fix and estimate that the true value is between 22\% and 38\% of lines. However, this is only true if all changes are considered. If only changes to production code files are considered, the average number of bug fixing lines is between 69\% and 93\%.  
\end{mdframed}

\subsubsection{Impact of Tangling}
\label{sec:impact}

The impact of tangling on software engineering research depends on the type of tangling as well as the research application. 
Moreover, tangling is not, by definition, bad. For example, adding new tests as part of a bug fix is actually a best practice. In the following, we use the term \emph{benign tangling} to refer to tangled commits that do not negatively affect research and \emph{problematic tangling} to refer to tangling that results in noise within the data without manual intervention. Specifically, we discuss problematic tangling for the three research topics we already mentioned in the motivation, i.e., program repair, bug localization, and defect prediction. 

If bug fixes are used for program repair, only changes to production code are relevant, i.e., all other tangling is benign as the tangling can easily be ignored using regular expressions, same as we did in this article. In production code files, the tangling of whitespaces and documentation changes is also benign. Refactorings and unrelated improvements are problematic, as they are not required for the bug fix and needlessly complicate the problem of program repair. If bug fixes are used as training data for a program repair approach~\citep[e.g.,][]{Li2020}, this may lead to models that are too complex as they would mix the repair with additional changes. If bug fixes are used to evaluate the correctness of automatically generated fixes~\citep[e.g.,][]{Martinez2016}, this comparison may be more difficult or noisy, due to the tangling. Unrelated improvements are especially problematic if they add new features. This sort of general program synthesis is usually out of scope of program repair and rather considered as a different problem, e.g., as neural machine translation~\citep{Tufano2019} and, therefore, introduces noise in dedicated program repair analysis tasks as described above.

For bug localization, tangling outside of production code files is also irrelevant, as these changes are ignored for the bug localization anyways. Within production code files, all tangling is irrelevant, as long as a single line in a file is bug fixing, assuming that the bug localization is at the file level.

For defect prediction, we assume that a state-of-the-art SZZ variant that accounts for whitespaces, documentations, and automatically detectable refactorings is used~\citep{Neto2018} to determine the bug fixing and bug inducing commits. Additionally, we assume that a heuristic is used to exclude non-production changes, which means that changes to non-production code files are not problematic for determining bug labels. For the bug fixes, we have the same situation as with bug localization: labeling of files would not be affected by tangling, as long as a single line is bug fixing. However, all unrelated improvements and non-automatically detectable refactorings may lead to the labeling of additional commits as inducing commits and are, therefore, potentially problematic. Finally, defect prediction features can also be affected by tangling. For example, the popular metrics by \cite{Kamei2013} for just-in-time defect prediction compute change metrics for commits as a whole, i.e., not based on changes to production code only, but rather based on all changes. Consequently, all tangling is problematic with respect to these metrics. According to the histogram for overall tangling in Figure~\ref{fig:results-commit-level}, this affects most commits. 

\begin{table}
\centering
\begin{tabular}{lrr}
\textbf{Research Topic} & \textbf{Bugs} & \textbf{File Changes} \\
\hline\hline
Program repair & 12\%--35\% & 9\%--32\% \\
Bug localization & 9\%--23\% & 7\%--21\% \\
Defect prediction (bug fix) & 3\%--23\% & 2\%--21\% \\
Defect prediction (inducing) & 5\%--24\% & 3\%--18\% \\
Defect prediction (total) & 8\%--47\% & 5\%--39\% \\
\hline
\end{tabular}
\caption{The ratio of problematic tangling within our data with respect to bugs and production file changes. The values for the inducing commits in defect prediction are only the commits that are affected in addition to the bug fixing commits.}
\label{tbl:problematic}
\end{table}

Table~\ref{tbl:problematic} summarizes the presence of problematic tangling within our data. We report ranges, due to the uncertainty caused by lines without consensus. The lower bound assumes that all lines without consensus are bug fixing and that all refactorings could be automatically detected, whereas the upper bound assumes that the lines without consensus are unrelated improvements and refactorings could not be automatically detected. We observe that all use cases are affected differently, because other types of tangled changes cause problems. Program repair has the highest lower bound, because any refactoring or unrelated improvement is problematic. Bug localization is less affected than defect prediction, because only the bug fixing commits are relevant and the bug inducing commits do not matter. If bug localization data sets would also adopt automated refactoring detection techniques, the numbers would be the same as for defect prediction bug fix labels. However, we want to note that the results from this article indicate that the automated detection of refactorings may be unreliable and could lead to wrong filtering of lines that actually contribute to the bug fix. Overall, we observe that the noise varies between applications and also between our best case and worst case assumptions. In the best case, we observe as little as 2\% problematically tangled bugs (file changes for bug fixes in defect prediction), whereas in the worst case this increases to 47\% (total number of bugs with noise for defect prediction). Since prior work already established that problematic tangling may lead to wrong performance estimations~\citep{Kochar2014, Herzig2016, Mills2020} as well as the degradation of performance of machine learning models~\citep{Nguyen2013}, this is a severe threat to the validity of past studies.  

\begin{mdframed}
Tangled commits are often \emph{problematic}, i.e., lead to noise within data sets that cannot be cleaned using heuristics. The amount of noise varies between research topics and is between an optimistic 2\% and a pessimistic value of 47\%. As researchers, we should be skeptical and assume that unvalidated data is likely very noisy and a severe threat to the validity of experiments, until proven otherwise. 
\end{mdframed}

\subsubsection{Reliability of Labels}

Participants have a probability of random mistakes of about 7.5\% overall and 9.0\% in production code files. Due to these numbers, we find that our approach to use multiple labels per commit was effective. Using the binomial distribution to determine the likelihood of at least three random mislabels, we expect that 111 lines with a consensus label are wrong, i.e., an expected error rate of 0.09\% in consensus labels. This error rate is lower than, e.g., using two people that would have to agree. In this case, the random error rate would grow to about 0.37\%. We fail to achieve consensus on 8.9\% of all lines and 14.3\% of lines in production code files. Our data indicates that most lines without consensus are not due to random mistakes, but rather due to disagreements between participants whether a line contributes to the bug fix or not. This indicates that the lines without consensus are indeed hard to label. Our participant survey supports our empirical analysis of the labeled data. A possible problem with the reliability of the work could also be participant dependent ``default behavior'' in case they were uncertain. For example, participants could have labeled lines as bug fixing in case they were not sure, which would reduce the reliability of our work and also possibly bias the results towards more bug fixing lines. However, we have no data regarding default behavior and believe that this should be studied in an independent and controlled setting. 

\begin{mdframed}
We reject H2 that participants fail to achieve consensus on at least 10.5\% of lines and find that this is not true when we consider all changes. However, we observe that this is the case for the labeling of production code files with 14.3\% of lines without consensus. Our data indicates that these lines are hard to label by researchers with active disagreement instead of random mistakes. Nevertheless, the results with consensus are reliable and should be close to the ground truth. 
\end{mdframed}

\subsubsection{Comparison to Prior Work}

Due to the differences in approaches, we cannot directly compare our results with those of prior studies that quantified tangling. However, the fine-grained labeling of our data allows us to restrict our results to mimic settings from prior work. From the description in their paper, \cite{Herzig2013a} flagged only commits as tangled, that contained source code changes for multiple issues or clean up of the code. Thus, we assume that they did not consider documentation changes or whitespace changes as tangling. This definition is similar to our definition of problematic tangling for program repair. The 12\%--35\% affected bugs that we estimate includes the lower bound of 15\% on tangled commits. Thus, our work seems to replicate the findings from \cite{Herzig2013a}, even though our methodologies are completely different. Whether the same holds true for the 22\% of tangled commits reported by \cite{Nguyen2013} is unclear, because they do not specify how they handle non-production code. Assuming they ignore non-production code, our results would replicate those by \cite{Nguyen2013} as well. 

\cite{Kochar2014} and \cite{Mills2020} both consider tangling for bug localization, but have conflicting results. Both studies defined tangling similar to our definition of problematic tangling for bug localization, except that they did not restrict the analysis to production code files, but rather to all Java files. When we use the same criteria, we have between 39\% and 48\% problematic tangling in Java file changes. Similar to what we discussed in Section~\ref{sec:impact}, the range is the result of assuming lines with consensus either as bug fixes or as unrelated improvements. Thus, our results replicate the work by \cite{Mills2020} who found between 32\% and 50\% tangled file changes. We could not confirm the results by \cite{Kochar2014} who found that 28\% of file changes are problematic for bug localization. 

Regarding the types of changes, we can only compare our work with the data from \cite{Nguyen2013} and \cite{Mills2020}. The percentages are hard to compare, due to the different levels of abstraction we consider. However, we can compare the trends in our data with those of prior work. The results regarding tangled test changes are similar to \cite{Mills2020}. For the other types of changes, the ratios we observe are more similar to the results reported by \cite{Mills2020} than those of \cite{Nguyen2013}. We observe only few unrelated improvements, while this is as common as tangled documentation changes in the work by \cite{Nguyen2013}. Similarly, \cite{Mills2020} observed a similar amount of refactoring as documentation changes. In our data, documentation changes are much more common than refactorings, both in all changes as well as in changes to production code files. A possible explanation for this are the differences in the methodology: refactorings and unrelated improvement labels are quite common in the lines without consensus. Thus, detecting such changes is common in the difficult lines. This could mean that we underestimate the number of tangled refactorings and unrelated improvements, because many of them are hidden in the lines without consensus. However, \cite{Nguyen2013} and \cite{Mills2020} may also overestimate the number of unrelated improvements and refactorings, because they had too few labelers to have the active disagreements we observed. Regarding the other contradictions between the trends observed by \cite{Nguyen2013} and \cite{Mills2020}, our results indicate that the findings by \cite{Mills2020} generalize to our data: we also find more refactorings than other unrelated improvements. 

\begin{mdframed}
Our results confirm prior studies by \cite{Herzig2013a}, \cite{Nguyen2013}, and \cite{Mills2020} regarding the prevalence of tangling. We find that the differences in estimations are due to the study designs, because the authors considered different types of tangling, which we identify as different types of problematic tangling. Our results for tangled change types are similar to the results by \cite{Mills2020}, but indicate also a disagreement regarding the prevalence of refactorings and unrelated improvements, likely caused by the uncertainty caused by lines that are difficult to label. 
\end{mdframed}

\subsubsection{Anectodal Evidence on Corner Cases}

Due to the scope of our analysis, we also identified some interesting corner cases that should be considered in future work on bug fixes. Sometimes, only tests were added as part of a bug fixing commit, i.e., no production code was changed. Usually, this indicates that the reported issue was indeed not a bug and that the developers added the tests to demonstrate that the case works as expected. In this case, we have a false positive bug fixing commit, due to a wrong issue label. However, sometimes it also happened, that the label was correct, but the bug fixing commit still only added tests. An example for this is the issue IO-466,\footnote{\url{https://issues.apache.org/jira/browse/IO-466}} where the bug was present, but was already fixed as a side effect of the correction of IO-423. This means we have an issue that is indeed a bug and a correct link from the version control system to the issue tracking system, but we still have a false positive for the bug fixing commit. This shows that even if the issue types and issue links are manually validated, there may be false positives in the bug fixing commits, if there is no check that production code is modified. 

Unfortunately, it is not as simple as that. We also found some cases, where the linked commit only contained the modification of the change log, but no change to production code files. An example for this is NET-270.\footnote{\url{https://issues.apache.org/jira/browse/NET-270}} This is an instance of missing links: the actual bug fix is in the parent of the linked commit, which did not reference NET-270. Again, we have a correct issue type, correct issue link, but no bug fix in the commit, because no production code file was changed. However, a directly related fix exists and can be identified through manual analysis. 

The question is how should we deal with such cases in future work? We do not have clear answers. Identifying that these commits do not contribute to the bug fix is trivial, because they do not change production code files. However, always discarding such commits as not bug fixing may remove valuable information, depending on the use case. For example, if we study test changes as part of bug fixing, the test case is still important. If you apply a manual correction, or possibly even identify a reliable heuristic, to find fixes in parent commits, the second case is still important. From our perspective, these examples show that even the best heuristics will always produce noise and that even with manual validations, there are not always clear decisions. 

\begin{figure}
\centering
\begin{minipage}[t]{0.3\textwidth}
Bug:
\begin{lstlisting}
public void foo() {
  a.foo();
}
\end{lstlisting}
\end{minipage}
$\quad$
\begin{minipage}[t]{0.3\textwidth}
Fix 1:
\begin{lstlisting}
public void foo() {
  if(a!=null) {
    a.foo();
  }
}
\end{lstlisting}
Diff for Fix 1:
\begin{lstlisting}
+   if(a!=null) {
-   a.foo()
+     a.foo()
+   }
\end{lstlisting}
\end{minipage}
$\quad$
\begin{minipage}[t]{0.3\textwidth}
Fix 2:
\begin{lstlisting}
public void foo() {
  if(a==null) {
    return;
  }
  a.foo();
}
\end{lstlisting}
Diff for Fix 2:
\begin{lstlisting}
+  if(a==null) {
+    return;
+  }
\end{lstlisting}
\end{minipage}
\caption{Example for a bug fix, where a new condition is added. In Fix 1, the line \texttt{a.foo()} is modified by adding whitespaces and part of the textual difference. In Fix 2, \texttt{a.foo()} is not part of the diff.}
\label{fig:whitespace-problem}
\end{figure}

An even bigger problem is that there are cases where it is hard to decide which modifications are part of the bug fix, even if you understand the complete logic of the correction. A common case of this is the ``new block problem'' that is depicted in Figure~\ref{fig:whitespace-problem}. The problem is obvious: the unchecked use of \texttt{a.foo()} can cause a \texttt{NullPointException}, the fix is the addition of a null-check. Depending on how the bug is fixed, \texttt{a.foo()} is either part of the diff, or not. This leads to the question: is moving \texttt{a.foo()} to the new block part of the bug fix or is this ``just a whitespace change''? The associated question is, whether the call to \texttt{a.foo()} is the bug or if the lack of the null-check is the bug. What are the implications that the line with \texttt{a.foo()} may once be labeled as part of the bug fix (Fix 1), and once not (Fix 2)? We checked in the data, and most participants labeled all lines in both fixes as contributing the bug fix in the cases that we found. There are important implications for heuristics here: 1) pure whitespace changes can still be part of the bug fix, if the whitespace changes indicates that the statement moved to a different block; 2) bugs can sometimes be fixed as modification (Fix 1), but also as pure addition (Fix 2), which leads to different data. The first implication is potentially severe for heuristics that ignore whitespace changes. In this case, textual differences may not be a suitable representation for reasoning about the changes and other approaches, such as differences in abstract syntax trees~\citep{Yang1991}, are better suited.
The second is contrary to the approach by \cite{Mills2020} for untangling for bug localization, i.e., removing all pure additions. In this case the addition could also be a modification, which means that the bug could be localized and that ignoring this addition would be a false negative. Similarly, the SZZ approach to identify the bug inducing commits cannot find inducing commits for pure additions. Hence, SZZ could blame the modification of the line \texttt{a.foo()} from Fix 1 to find the bug inducing commit, which would be impossible for Fix 2.

\subsubsection{Implications for Researchers}

While we have discussed many interesting insights above, two aspects stand out and are, to our mind, vital for future work that deals with bugs. 

\begin{itemize}
    \item \textit{Heuristics are effective!} Most tangling can be automatically identified by identifying non-production code files (e.g., tests), and changes to whitespaces and documentation within production code files.\footnote{For example, the pycoSHARK contains methods for checking if code is Java Production code for our projects and all standard paths for tests and documentation. The VisualSHARK and the inducingSHARK are both able to find whitespace and comment-only changes. All tools can be found on GitHub: \url{https://github.com/smartshark}} Any analysis that should target production code but does not carefully remove tests cannot be trusted, because test changes are more common than bug fixing changes. 
    \item \textit{Heuristics are imperfect!} Depending on the use case and the uncertainty in our data, up to 47\% could still be affected by tangling, regardless of the heuristics used to clean the data. We suggest that researchers should carefully assess which types of tangling are problematic for their work and use our data to assess how much problematic tangling they could expect. Depending on this estimation, they could either estimate the threat to the validity of their work or plan other means to minimize the impact of tangling on their work.
\end{itemize}

\subsubsection{Summary for RQ1}

In summary, we have the following result for RQ1.

\begin{mdframed}
We estimate that only between 22\% and 38\% of changed lines within bug fixing commits contribute to the functional correction of the code. However, much of this additional effort seems to be focused on changes to non-production code, like tests or documentation files, which is to be expected in bug fixing commits. For production code, we estimate that between 69\% and 93\% of the changes contribute to the bug fix. We further found that researchers are able to reliably label most data, but that multiple raters should be used as there is otherwise a high likelihood of noise within the data.
\end{mdframed}

\section{Experiments for RQ2}
\label{sec:rq2}

We now present our results and discuss the implications for RQ2 on the effect of gamification for our study.

\subsection{Results for RQ2}
\label{sec:participation}

\begin{figure}
\centering
\includegraphics[width=\textwidth]{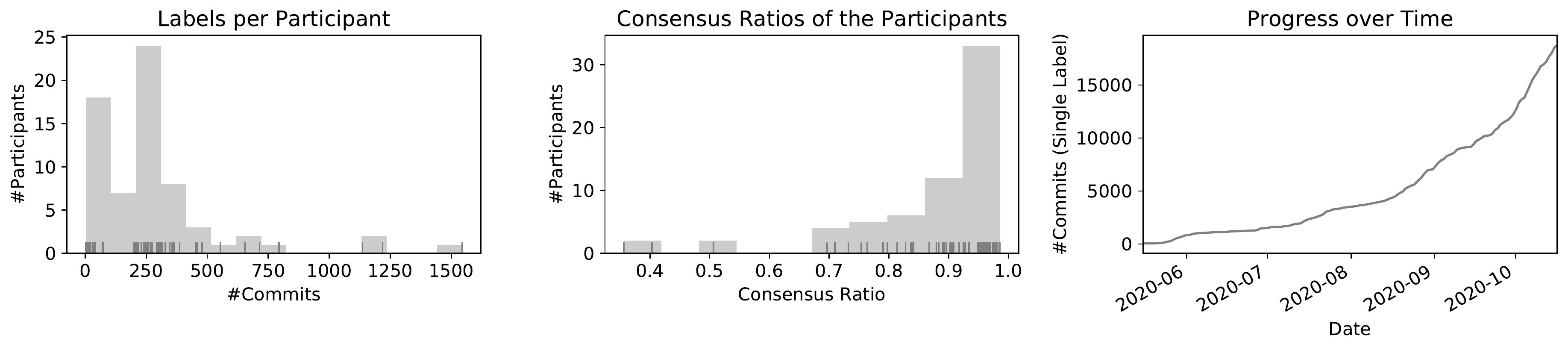}
\caption{Number of commits labeled per participant and number of commit labels over time.}
\label{fig:participant_stats}
\end{figure}

\begin{figure}
\centering
\includegraphics[width=\textwidth]{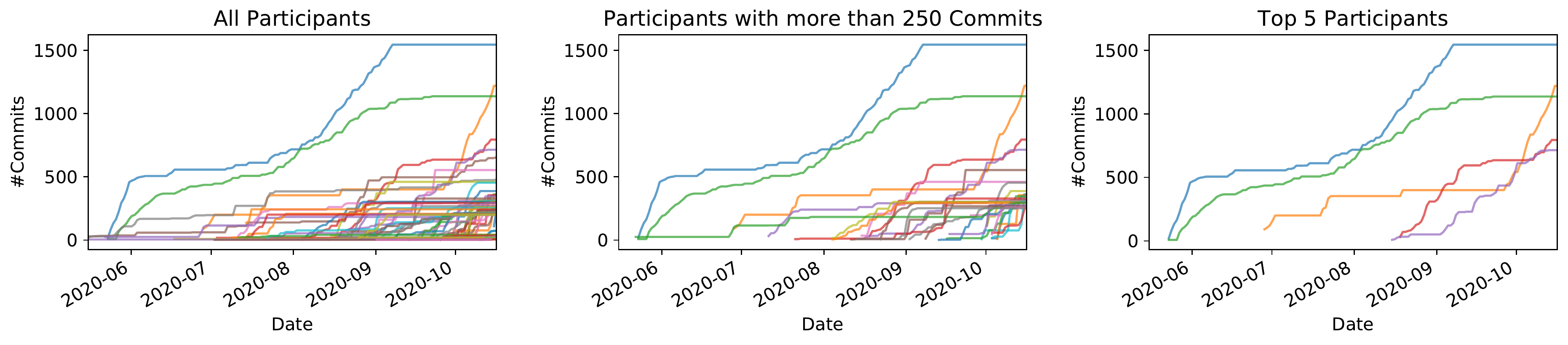}
\caption{Lines labeled per participants over time. Each line is one participant. The plot on the left shows the data for all participants. The plot in the middle shows only participants who labeled more than 250 commits, excluding Steffen Herbold, Alexander Trautsch, and Benjamin Ledel because their position in the author list is not affected by the number of labeled commits. The plot on the right shows only the top 5 participants.}
\label{fig:progress_over_time}
\end{figure}

Figure~\ref{fig:participant_stats} shows how many commits were labeled by each participant and how labeling progressed over time. We observe that most participants labeled around 200 commits and that the labeling started slowly and accelerated at the end of August.\footnote{Details about when we recruited participants are provided in Appendix~\ref{sec:recruitment}.} Moreover, all participants with more than 200 commits achieved at least 70\% consensus and most participants were clearly above this lower boundary. Five participants that each labeled only few lines before dropping out are below 70\%. We manually checked, and found that the low ratio was driven by single mistakes and that each of these participants have participated in the labeling of less than 243 lines with consensus. This is in line with our substantial agreement measured with Fleiss' $\kappa$. 

Figure~\ref{fig:progress_over_time} shows how many lines were labeled by each participant over time. The plot on the left confirms that most participants started labeling data relatively late in the study. The plot in the middle is restricted to the participants that have labeled substantially more lines than required, which resulted in being mentioned earlier in the list of authors. We observe that many participants are relatively densely located in the area around 250 commits. These participants did not stop at 200 commits, but continued a bit longer before stopping. Seven participants stopped when they finished 200 bugs, i.e., they mixed up the columns for bugs finished with commits finished in the leaderboard, which may explain some of the data. We could see no indication that these participants stopped because they achieved a certain rank. The plot on the right shows the top five participants that labeled most data. We observe that the first and third ranked participant both joined early and consistently labeled data over time. We note that the activity of the first ranked participant decreased, until the third ranked participant got close, and then accelerated again. The second ranked participant had two active phases of labeling, both of which have over 500 labeled commits. From the data, we believe that achieving the second rank over all may have been the motivation for labeling here. The most obvious example of the potential impact of the gamification on activity are the fourth and fifth ranked participants. The fourth ranked stopped labeling, until they were overtaken by the fifth ranked participant, then both changed places a couple of times, before the fourth ranked participant prevailed and stayed on the fourth rank. Overall, we believe that the line plot indicates these five participants were motivated by the gamification to label substantially more. Overall, these five participants labeled 5,467 commits, i.e., produced 5.4 times more data than was minimally required by them.

\begin{figure}
\centering
\includegraphics[width=0.5\textwidth]{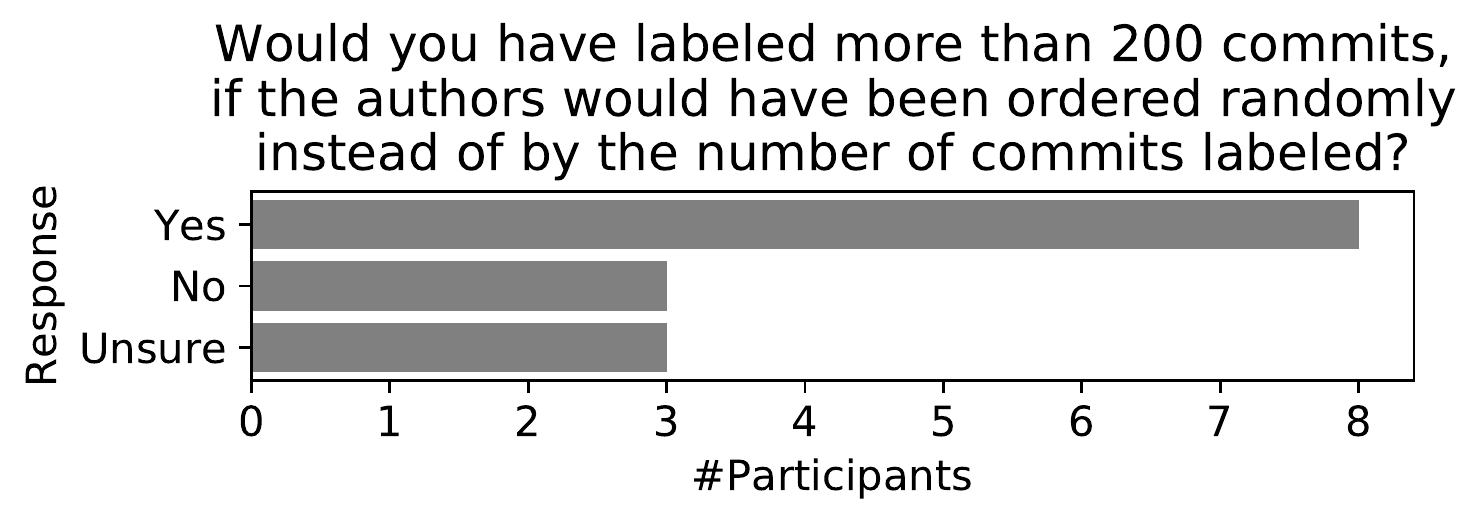}
\caption{Histogram of the answers to the question Q2: Would you have labeled more than 200 commits, if the authors would have been ordered randomly instead of by the number of commits labeled? 14 participants out of 26 with at least 250 commits answered this question.}
\label{fig:author-order}
\end{figure}

Figure~\ref{fig:author-order} shows the results of our question if participants with at least 250 commits would have labeled more than 200 commits, if the author order would have been random. The survey indicates that most participants did not care about the gamification element, which is in line with the label data we observed. However, we also see that three participants answered with ``No'', i.e., they were definitively motivated by the gamification and would have labeled less data otherwise. Another three answered with ``Unsure'', i.e., they were also motivated by the gamification, but would possibly still have labeled the same amount of data otherwise. Thus, the gamification increased the amount of effort invested in this study for three to six participants, i.e.,  participants were motivated by the gamification. This means that between 7\% and 13\% of the recruited participants\footnote{Not counting the three principal investigators Steffen Herbold, Alexander Trautsch and Benjamin Ledel} who fulfilled the criteria for co-authorship were motivated by the gamification. 

\subsection{Discussion of RQ2}

Our data yields relatively clear results. While only between 7\% and 13\% of participants labeled more data due to the gamification element, the impact of this was still strong. The five participants that invested the most effort labeled 5,467 commits, which is roughly the same amount of data as the 24 participants who labeled at least 200 commits and are closest to 200 commits. A risk of this gamification is that this could negatively affect the reliability of our results, e.g., the participants who labeled most data would be sloppy. However, this was not the case, i.e., the consensus ratios are consistently high and, hence, independent of the amount of commits that were labeled. While we have no data to support this, our requirements of a minimal consensus ratio for co-authors may have prevented a potential negative impact of the gamification, because the penalty for sloppiness would have been severe. In summary, we have the following results for RQ2.

\begin{mdframed}
We fail to reject H3 and believe that motivating researchers to invest more effort through gamification is worthwhile, even if only a minority is motivated.
\end{mdframed}

\section{Threats to Validity}
\label{sec:threats}

We report the threats to the validity of our work following the classification by \cite{Cook1979} suggested for software engineering by \cite{Wohlin2012}. Additionally, we discuss the reliability as suggested by \cite{Runeson2009}. 

\subsection{Construct Validity}

There are several threats to the validity of our research construct. We differentiate not only between bug fixing lines and other lines, but use a more detailed differentiation of changes into whitespaces only, documentation changes, test changes, refactorings, and unrelated improvements. This could have influenced our results. Most notably, this could have inflated the number of lines without consensus. We counter this threat by differentiating between lines without consensus with at least one bug fix label, and other lines without consensus. 

Another potential problem is our assumption that minority votes are random mistakes and that random mistakes follow a binomial distribution. However, we believe that unless the mistakes follow a binomial distribution, i.e., are the result of repeated independent Bernoulli experiments, they are not truly random. Our data indicates that there is support for this hypothesis, i.e., that there are also many active disagreements, but that these are unlikely to be in cases where we have consensus.

We also ignore a potential learning curve of participants for our evaluation of mistakes and, therefore, cannot rule out that the amount of mistakes decreases over time. We mitigate these issues through the five tutorial bugs that all participants had to label at the beginning, which provides not only a training for the tool usage, but possibly also flattens the learning curve for the bug labeling. 

Furthermore, the participants may have accepted pre-labeled lines without checking if the heuristic pre-labeling was correct. We mitigated this bias by advising all participants to be skeptical of the pre-labels and carefully check them, as they are only hints. The disagreements with the refactoring pre-labels indicate that participants were indeed skeptical and did not accept pre-labeled lines without checking. While all projects use Java as the main programming language, some projects also use additional languages, which means we may slightly underestimate the amount of production code.

The results regarding the gamification may be untrustworthy, because the participants were fully aware that this would be analyzed. Thus, a participant could have, e.g., avoided to label more data, if they wanted to show that gamification is ineffective, or labeled more data, if they wanted to show that gamification is effective. We have no way to counter this threat, since not revealing this aspect of the analysis as part of the pre-registered study protocol would not have been in line with our open science approach\footnote{Pre-registration, all tools are open source, all data publicly available.} and be ethically questionable. Instead, we rely on our participants, who are mostly researchers, to act ethical and not intentionally modify our data through such actions. Additionally, one may argue that authorship order is not really gamification, because the reward is not a game element, but rather an impact on the real world. Gamification elements that would have no tanglible real world benefit, e.g., badges that could be earned while labeling, may lead to different results. 

Similarly, the participants who answered the questions regarding the estimation of the lines where they were unsure or if they would have labeled more than 200 commits if the author order would have been randomized were effectively answered by the authors of this article. Thus, we could have constructed the results in a way to support our findings. Again, we can only rely on the ethical behavior of all participants but want to note that, in the end, any anonymous survey has similar risks. Moreover, we see no benefit for anybody from answering the questions incorrectly. 

Finally, there were several opportunities for additional surveys among the participants, e.g., to directly ask specific participants if the gamification was a motivating factor or if they followed the provided guidance in specific aspects. We decided against this because this was not part of the pre-registered protocol and such questions could put participants into ethically problematic situations, e.g., because they may feel that stating that they were driven by curiosity instead of benefit would be beneficial. This is just hypothetical, but we still want to avoid such situations. 

\subsection{Conclusion Validity}

There are no threats to the conclusion validity of our study. All statistical tests are suitable for our data, we correct our results to err on the side of caution, and the pre-registration ensured that we did not tailor our statistical analysis in a way to articifially find coincidental results, e.g., through subgroup analysis. 

\subsection{Internal Validity}

\begin{figure}
\centering
\includegraphics[width=\textwidth]{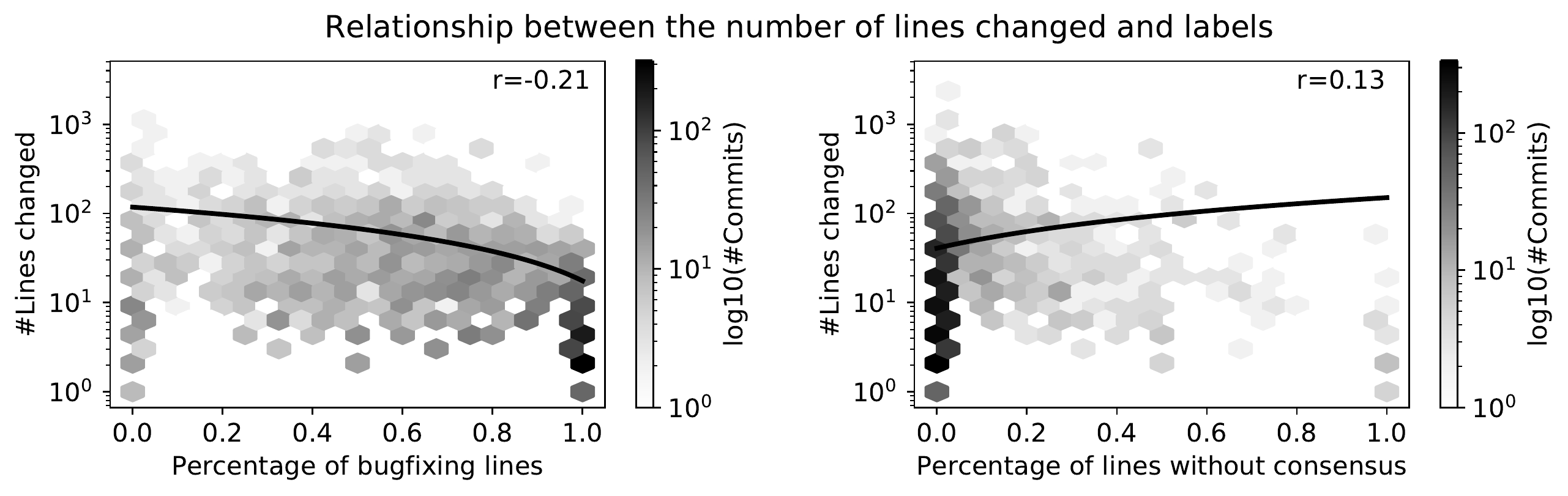}
\includegraphics[width=\textwidth]{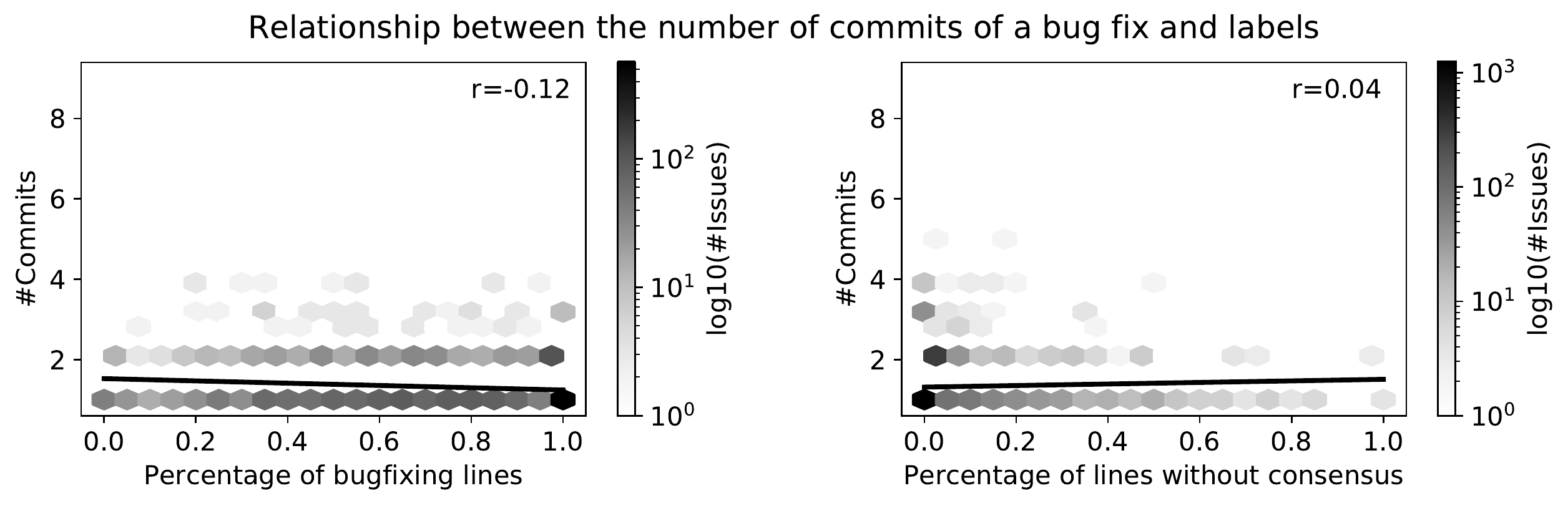}
\caption{Relationship between the number of lines changed in production code files and the percentage of bug fixing lines and lines without consensus. The line indicates the regression line for a linear relationship between the variables, the $r$-value is Pearson's correlation coefficient~\cite{Kirch2008}. The correlations are significant with $p<0.01$ for all values. The regression line has a curve in the upper plots, due to the logarithmic scale of the y-axis.}
\label{fig:impact-of-size}
\end{figure}

The main threat to the internal validity of our work is that our analysis of percentages of change types per commit could be misleading, because the distribution of the percentages is not mainly driven by tangling, but by a different factor. The only such factor that we consider to be a major threat is the size: larger commits could be more difficult to label and, hence, have a different distribution of mislabels. Moreover, larger commits could indicate that not only the bug was fixed, but that there are many tangled changes. Similar arguments can be used for the number of commits used to fix a bug. Figure~\ref{fig:impact-of-size} explores the relationship between the number of changed lines, as well as the number of commits and the percentage of bug fixing lines, respectively the lines without consensus. Both plots reveal that there is no pattern within the data that is related to the size and that no clear correlation is visible. We find very weak linear correlations that are significant between the number of lines changed and the ratio of bug fixing lines. However, given the structure of the data, this rather looks like a random effect than a clear pattern. The significance is not relevant in our case, as even weak correlations are significant given our amount of data. Another factor could also be that our results are project dependent. However, since this would be a case of subgroup analysis, we cannot simply extend our study protocol to evaluate this without adding a risk of finding random effects. Moreover, adding this analysis would invalidate one major aspect of pre-registrations, which is to avoid unplanned subgroup analysis. 

Our decision to label data on a line level and not the character or statement level may affect the validity of our conclusions. In case changes are tangled within the same line, this is not detected by our labeling, since we advised participants to label lines as bug fixing in that case.\footnote{Within the FAQ on \url{https://smartshark.github.io/msr20\_registered\_report/}} However, such changes are very rare, as this requires not only lines with multiple concerns, but also that these different concerns are modified within the same commit, such that at least one modification is not part of the bug fix. While we do not have data on how often exactly that was the case, our participants reported that this happened only rarely. Thus, any noise we missed this way is likely to have only a negligible effect on our results.

Another threat to the internal validity is that we may underestimate the ability of researchers to achieve consensus on tangling. More training, e.g., on how to use the labeling tool or how to deal with special cases could improve the consensus among participants in future studies. Moreover, our study design does not include a disagreement resolution phase, where participants can discuss to resolve disagreements. We opted against such a phase due to the scale of our analysis, but hope that one of the aspects of future work is to gain further insights into disagreements, including their potential resolution. 

Participants may have missed the gamification aspect in the study protocol and, hence, were not aware of how the author ranking would be determined. This could affect our conclusions regarding the impact of the gamification. However, missing the gamification aspect should only lead to underestimating the impact of the gamification, which means that our results should at least be reliable as a lower boundary. Moreover, participants may have labeled more than 200 commits, because the leaderboard was only updated once a day and they did not notice they passed 200 commits until the next day. However, our data indicates that if this was the case, this only led to a small amount of additional effort, i.e., at most 50 commits. 

\subsection{External Validity}

Due to our restriction to mature Java software, our results may not generalize to other languages or software with less mature development processes than those of the Apache projects. Moreover, the sample of projects may not be representative for mature Java projects as well, as we did not conduct random sampling, but re-used data from \cite{Herbold2019} who used a purposive sampling approach~\citep{Patton2014}. However, \cite{Baltes2020} argue that well-done purposive sampling should also yield generalizable results. Moreover, our analysis was done on open source projects. While there are many industrial contributors in the projects we studied, our results may not generalize to closed source software. 

\subsection{Reliability}

We cannot rule out that the results of our study are due to the participants we recruited as fellow researchers. However, our participants are very diverse: they are working on all continents except Africa and Antarctica, the experience ranges from graduate students to full professors, and from a couple of years of development to several decades. Regardless of this, the overall consensus ratios are relatively high for all participants and Fleiss' $\kappa$ indicates substantial agreement among our participants. 

\section{Conclusion}
\label{sec:conclusion}

We now conclude our article with a short summary and an outlook on possible future work on the topic.

\subsection{Summary}

Our study shows that tangled commits have a high prevalence. While we found that most tangling is benign and can be identified by current heuristics, there is also problematic tangling that could lead to up to 47\% of noise in data sets without manual intervention. We found that the identification of tangled commits is possible in general, but also that random errors have to be accounted for through multiple labels. We also found that there are about 14\% of lines in production code files where participants actively disagree with each other, which we see as an indication that these lines are especially difficult to label. Overall, our study showed that tangling is a serious threat for research on bugs based on repository mining that should always be considered, at least to assess the noise which may be caused by the tangling. 

\subsection{Future Work}

There are many opportunities for future work, many of them enabled directly by the data we generated: the data can be used as accurate data for program repair, bug localization, defect prediction, but also other aspects like the assessment of the relation between static analysis warnings or technical debt and bugs. Moreover, many studies were conducted with tangled data. If and how this affected the research results requires further attention. Based on prior work by \cite{Herzig2016} and \cite{Mills2020} there are already indications that tangling may have a significant impact on results. These are just some examples of use cases for this data and we fully expect the community to have many more good ideas. 

Moreover, while we have advanced our understanding of tangled commits, there are still important issues which we do not yet understand and that are out of scope of our study. Some of these aspects can be analyzed with the help of our data. For example, lines without consensus could undergo further inspection to understand why no consensus was achieved. Through this, we may not only get a better understanding of tangling, but possibly also about program comprehension. For example, small bug fixes without consensus could be an indicator that even small changes may be hard to understand, which in turn could indicate that this may be because the concept of atoms of confusion does not only apply to C/C++ \citep{Gopstein2018}, but also to Java. Benign tangling could also undergo further scrutiny: how many tangled documentation or test changes are directly related to the bug fix and how many are completely unrelated. This could also be correlated with other factors, e.g., to understand when code documentation must be updated as part of bug fixes to avoid outdated documentation. 

However, there are also aspects of tangling, which should not or cannot be studied on our data. For example, how the prevalence of tangling commits varies between projects and which factors influence the degree of tangling in a project. While our data could certainly be used to derive hypothesis for this, independent data must be used to evaluate these hypotheses as this would be a case of post-hoc subgroup analysis that could lead to overinterpretation of results that are actually non-reproducible random effects. Another gap in our knowledge is tangling beyond bug fixing commits, for which we still only have limited knowledge. Moreover, while we do not see any reason why tangling should be different in other programming languages, we also have no data to support that our results generalize beyond Java. Similarly, it is unclear how our results translate to closed source development, that possibly follows stricter rules regarding the use of issue tracking systems and the implementation of changes, which could reduce the tangling. 

Finally, we note that the research turk was a suitable research method for this study and we look forward to future studies that follow a similar approach. Since even the very basic gamification approach we used was effective, future studies should consider to use more game elements (e.g., badges) and evaluate if they are effective at increasing the participation. 

\section*{Acknowledgments}

Alexander Trautsch and Benjamin Ledel and the development of the infrastructure required for this research project were funded by DFG Grant 402774445. Ivan Pashchenko was partially funded by the H2020 AssureMOSS project (Grant No. 952647). We thank the cloud team of the GWDG (\url{https://gwdg.de}) that helped us migrate the VisualSHARK to a stronger machine within a day, to meet the demands of this project once many participants started to label at the same time. 

\section*{Author Contributions}

Conceptualization: Steffen Herbold;
Methodology: Steffen Herbold, Alexander Trautsch, Benjamin Ledel;
Formal analysis and investigation: \textit{all authors};
Writing - original draft preparation: Steffen Herbold;
Writing - review and editing: \textit{all authors};
Funding acquisition: Steffen Herbold;
Resources: Steffen Herbold, Alexander Trautsch, Benjamin Ledel;
Supervision: Steffen Herbold;
Project administration: Steffen Herbold, Alexander Trautsch;
Software: Steffen Herbold, Alexander Trautsch, Benjamin Ledel.

\bibliography{./literature}

\appendix

\section*{Appendix}

\section{Email with Instructions}
\label{app:instructions}

Dear FIRST\_NAME,

\vspace{6pt}
\noindent
Thank you for joining us! You will get your account soon (at most a couple of days, usually a lot faster) from Alexander via Email. 

\vspace{6pt}
\noindent
Once you have your credentials, you can access the labeling system (1).

\vspace{6pt}
\noindent
Some more things, that are not in the tutorial video, but also important:
\begin{itemize}
\item The first five bugs that you are shown are scripted, i.e., everybody will have to work on these five bugs. They overlap with the tutorial. This should help you get comfortable with the system and avoid mislabels that could happen while get used to the system. 
\item We use the RefactoringMiner (2) to automatically label lines that are potentially refactorings to help you identify such lines. Please check this carefully. We already found several instances, where RefactoringMiner was wrong, e.g., because it overlooked side effects due to method calls. 
\item You can select the project you want to work on in the dropdown menu on the right side of the page. You may have to open the menu first by clicking on the three bars in the top-right corner.
\item If you are not sure which project you want to work on, check the leaderboard. The leaderboard not only lists how much the other participants have labeled, but also the progress for the individual projects. Ideally, work on projects where others have also started to label issues. 
\item If no bugs are shown for the project you selected, the reason could also be that the labeling for this project is already finished. You can check this in the leaderboard. 
\item While we manually validated if the issues are really bugs and the links between the issues and the commits, there is always a chance that we made a mistake or missed something. If you think that there is a wrong link to a bug fixing commit or that an issue is not really a bug, write me an email with the bug id (e.g., IO-111) and the problem you found. We will then check this and possibly correct this in the database. All such corrections will be recorded. 
\item Finally, if you find any bugs in our system, either file an issue on GitHub (3) or just write me an email. 
\end{itemize}

\vspace{6pt}
\noindent
Best,

\vspace{6pt}
\noindent
Steffen

\vspace{6pt}
\noindent
(1) https://visualshark.informatik.uni-goettingen.de/

\noindent
(2) https://github.com/tsantalis/RefactoringMiner

\noindent
(3) https://github.com/smartshark/visualSHARK

\section{Detailed Results for Lines without Consensus}

\begin{table}[h]
\centering
\begin{tabular}{lrrrrrr}
& Bug fix & Doc. & Refactoring & Unrelated & Whitespace & Test \\
\hline
Bug fix & - & $0.1 \pm 0.0$ & $3.6 \pm 0.0$ & $4.2 \pm 0.0$ & $0.7 \pm 0.0$ & $0.2 \pm 0.0$ \\
Documentation & $3.3 \pm 0.0$ & - & $0.8 \pm 0.0$ & $1.6 \pm 0.0$ & $0.9 \pm 0.0$ & $0.3 \pm 0.0$ \\
Refactoring & $13.7 \pm 0.0$ & $0.7 \pm 0.0$ & - & $3.4 \pm 0.0$ & $0.9 \pm 0.0$ & $0.0 \pm 0.0$ \\
Unrelated & $18.5 \pm 0.0$ & $0.8 \pm 0.0$ & $1.9 \pm 0.0$ & - & $1.2 \pm 0.0$ & $0.0 \pm 0.0$ \\
Whitespace & $4.2 \pm 0.0$ & $0.3 \pm 0.0$ & $2.8 \pm 0.0$ & $2.4 \pm 0.0$ & - & $0.5 \pm 0.0$ \\
\hline
\end{tabular}
\caption{Estimated probabilities of random mislabels in production code files. The rows are the correct labels, the columns are the mislabels.}
\label{tbl:prob-mislabels}
\end{table}

\begin{longtable}{lll}
\caption{Observed label combinations for lines without consensus.} \label{tbl:no-cons-lines} \\

\textbf{Observed Labels} & \textbf{\#Lines} & \textbf{Percentage} \\ \hline\hline
\endfirsthead

\multicolumn{3}{c}%
{{\bfseries \tablename\ \thetable{} -- continued from previous page}} \\
\hline 
\textbf{Observed Labels} & \textbf{\#Lines} & \textbf{Percentage} \\ \hline\hline
\endhead

\hline \multicolumn{3}{r}{{Continued on next page}} \\ \hline
\endfoot

\hline \hline
\endlastfoot

2 bug fix, 2 refactoring                            &  3680 &    14.6\% \\
2 bug fix, 2 unrelated                              &  3516 &    14.0\% \\
1 refactoring, 1 unrelated, 2 bug fix               &  1971 &     7.8\% \\
1 bug fix, 1 whitespace, 2 unrelated                &  1245 &     5.0\% \\
1 bug fix, 1 unrelated, 2 documentation             &  1130 &     4.5\% \\
1 bug fix, 1 documentation, 2 unrelated             &  1114 &     4.4\% \\
1 bug fix, 1 refactoring, 2 unrelated               &  1028 &     4.1\% \\
1 unrelated, 1 whitespace, 2 bug fix                &   902 &     3.6\% \\
2 bug fix, 2 documentation                          &   811 &     3.2\% \\
2 refactoring, 2 whitespace                        &   767 &     3.0\% \\
1 bug fix, 1 unrelated, 2 whitespace                &   719 &     2.9\% \\
2 bug fix, 2 whitespace                             &   688 &     2.7\% \\
1 bug fix, 1 unrelated, 2 refactoring               &   655 &     2.6\% \\
2 documentation, 2 refactoring                     &   633 &     2.5\% \\
2 documentation, 2 unrelated                       &   537 &     2.1\% \\
2 unrelated, 2 whitespace                          &   494 &     2.0\% \\
1 bug fix, 1 refactoring, 1 unrelated, 1 whitespace &   490 &     1.9\% \\
1 refactoring, 1 whitespace, 2 bug fix              &   489 &     1.9\% \\
1 documentation, 1 unrelated, 2 bug fix             &   435 &     1.7\% \\
1 bug fix, 1 whitespace, 2 refactoring              &   415 &     1.7\% \\
1 refactoring, 1 unrelated, 2 whitespace           &   402 &     1.6\% \\
1 bug fix, 1 refactoring, 2 whitespace              &   355 &     1.4\% \\
1 refactoring, 1 unrelated, 2 documentation        &   297 &     1.2\% \\
1 bug fix, 1 refactoring, 2 documentation           &   175 &     0.7\% \\
1 bug fix, 1 documentation, 2 refactoring           &   166 &     0.7\% \\
1 documentation, 1 refactoring, 2 whitespace       &   156 &     0.6\% \\
1 bug fix, 1 documentation, 1 refactoring, 1 unrelated &   148 &     0.6\% \\
2 refactoring, 2 unrelated                         &   140 &     0.6\% \\
1 refactoring, 1 test, 2 whitespace                &   139 &     0.6\% \\
1 documentation, 1 refactoring, 2 bug fix           &   129 &     0.5\% \\
1 unrelated, 1 whitespace, 2 refactoring           &   116 &     0.5\% \\
1 refactoring, 1 whitespace, 2 unrelated           &    97 &     0.4\% \\
2 documentation, 2 whitespace                      &    95 &     0.4\% \\
1 refactoring, 1 whitespace, 2 documentation       &    92 &     0.4\% \\
1 documentation, 1 test, 2 whitespace              &    89 &     0.4\% \\
2 bug fix, 2 test                                   &    85 &     0.3\% \\
1 test, 1 unrelated, 2 bug fix                      &    77 &     0.3\% \\
2 documentation, 2 test                            &    69 &     0.3\% \\
1 documentation, 1 unrelated, 2 refactoring        &    65 &     0.3\% \\
1 bug fix, 1 test, 2 unrelated                      &    61 &     0.2\% \\
1 test, 1 whitespace, 2 documentation              &    44 &     0.2\% \\
1 unrelated, 1 whitespace, 2 documentation         &    43 &     0.2\% \\
1 refactoring, 1 test, 2 documentation             &    31 &     0.1\% \\
1 bug fix, 1 documentation, 1 unrelated, 1 white... &    30 &     0.1\% \\
1 test, 1 whitespace, 2 bug fix                     &    29 &     0.1\% \\
2 refactoring, 2 test                              &    26 &     0.1\% \\
1 bug fix, 1 refactoring, 2 test                    &    26 &     0.1\% \\
1 bug fix, 1 documentation, 1 refactoring, 1 whitespace &    26 &     0.1\% \\
1 documentation, 1 whitespace, 2 refactoring       &    23 &     0.1\% \\
1 documentation, 1 whitespace, 2 bug fix            &    19 &     0.1\% \\
1 refactoring, 1 unrelated, 2 test                 &    16 &     0.1\% \\
1 bug fix, 1 refactoring, 1 test, 1 unrelated       &    16 &     0.1\% \\
1 bug fix, 1 documentation, 2 whitespace            &    15 &     0.1\% \\
1 bug fix, 1 whitespace, 2 test                     &    15 &     0.1\% \\
1 test, 1 unrelated, 2 whitespace                  &    12 &     $<$0.1\% \\
1 bug fix, 1 test, 2 documentation                  &    12 &     $<$0.1\% \\
1 bug fix, 1 test, 2 refactoring                    &    10 &     $<$0.1\% \\
1 bug fix, 1 test, 2 whitespace                     &    10 &     $<$0.1\% \\
1 bug fix, 1 whitespace, 2 documentation            &    10 &     $<$0.1\% \\
1 documentation, 1 refactoring, 2 unrelated        &     9 &     $<$0.1\% \\
1 test, 1 unrelated, 2 documentation               &     8 &     $<$0.1\% \\
1 documentation, 1 test, 2 unrelated               &     7 &     $<$0.1\% \\
1 bug fix, 1 unrelated, 2 test                      &     7 &     $<$0.1\% \\
2 test, 2 whitespace                               &     6 &     $<$0.1\% \\
1 refactoring, 1 test, 2 bug fix                    &     6 &     $<$0.1\% \\
1 documentation, 1 refactoring, 1 unrelated, 1 whitespace &     5 &     $<$0.1\% \\
1 documentation, 1 unrelated, 2 whitespace         &     4 &     $<$0.1\% \\
1 test, 1 whitespace, 2 refactoring                &     4 &     $<$0.1\% \\
1 bug fix, 1 test, 1 unrelated, 1 whitespace        &     3 &     $<$0.1\% \\
1 refactoring, 1 test, 2 unrelated                 &     3 &     $<$0.1\% \\
1 bug fix, 1 documentation, 2 test                  &     2 &     $<$0.1\% \\
1 bug fix, 1 documentation, 1 test, 1 unrelated     &     1 &     $<$0.1\% \\
1 bug fix, 1 documentation, 1 test, 1 whitespace    &     1 &     $<$0.1\% \\
\end{longtable}

\section{Recruitment of Participants}
\label{sec:recruitment}

\begin{figure}
\centering
\includegraphics[width=0.5\textwidth]{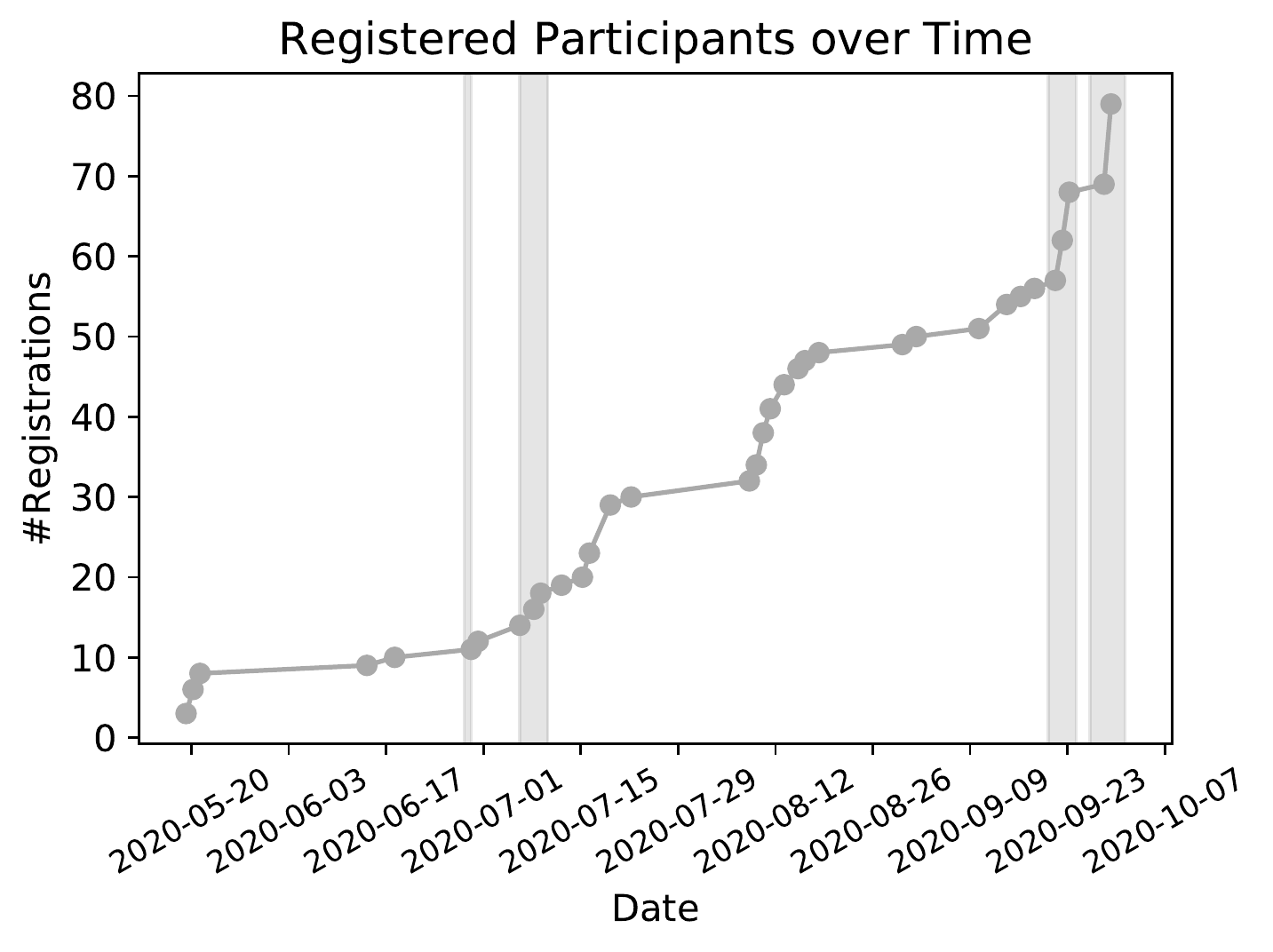}
\caption{Number of participants over time. The shaded areas show the time at which major software engineering conferences, during which we advertised for this project, took place. From left to right these dates are the MSR, the ICSE, the ASE, and the SCAM/ICSME.}
\label{fig:participants-over-time}
\end{figure}

Figure~\ref{fig:participants-over-time} shows that the number of registered participants increased throughout the timespan of the data collection phase. At the beginning, the increase was relatively slow, but recruitment picked up once we presented the registered report at the MSR 2020\footnote{\url{https://2020.msrconf.org/}} and the idea about the research turk at the ICSE 2020\footnote{\url{https://conf.researchr.org/home/icse-2020}}. Moreover, we actively advertised for the study at both conferences in the provided virtual communication media, which also helped to attract participants. Next, we see a sharp rise in the number of participants at the beginning of August. This is likely due to an announcement of the study on Twitter that we combined with asking all participants registered so far to share this and also to invite colleagues. The increase before the ASE 2020\footnote{\url{https://conf.researchr.org/home/ase-2020}} is the result of advertising for this study in software engineering related Facebook groups. The rise during the ASE is partially attributed to our advertisement at the ASE, but also due to many participants from a large research group joining at the same time, independent of the ASE. A final push for more participants at the SCAM and ICSME 2020\footnote{\url{https://icsme2020.github.io/}} was also effective and resulted in additional registrations. 

\end{document}